%% file: main.tex
\documentclass[prx, amsfonts, amssymb, amsmath, showkeys, superscriptaddress, twocolumn]{revtex4-1}

\input{preamble}

\begin{document}
\title{Noise-Assisted Variational Quantum Thermalization}

\author{Jonathan Foldager}
 \email[Correspondence: ]{jonf@dtu.dk}
 \affiliation{Section for Cognitive Systems, Department of Applied Mathematics and Computer Science, Technical University of Denmark (DTU), Kgs. Lyngby, Denmark}

\author{Arthur Pesah}
 \affiliation{Department of Physics and Astronomy, University College London, London WC1E 6BT, UK}
 
\author{Lars Kai Hansen}
\affiliation{Section for Cognitive Systems, Department of Applied Mathematics and Computer Science, Technical University of Denmark (DTU), Kgs. Lyngby, Denmark}

\begin{abstract}
Preparing thermal states on a quantum computer can have a variety of applications, from simulating many-body quantum systems to training machine learning models. Variational circuits have been proposed for this task on near-term quantum computers, but several challenges remain, such as finding a scalable cost-function, avoiding the need of purification, and mitigating noise effects. We propose a new algorithm for thermal state preparation that tackles those three challenges by exploiting the noise of quantum circuits. We consider a variational architecture containing a depolarizing channel after each unitary layer, with the ability to directly control the level of noise. We derive a closed-form approximation for the free-energy of such circuit and use it as a cost function for our variational algorithm. By evaluating our method on a variety of Hamiltonians and system sizes, we find several systems for which the thermal state can be approximated with a high fidelity. However, we also show that the ability for our algorithm to learn the thermal state strongly depends on the temperature: while a high fidelity can be obtained for high and low temperatures, we identify a specific range for which the problem becomes more challenging. We hope that this first study on noise-assisted thermal state preparation will inspire future research on exploiting noise in variational algorithms. 
\end{abstract}

\maketitle

\input{sections/1introduction}
\input{sections/2relatedwork}
\input{sections/3navqt}

\input{sections/4method}

\input{sections/5results}

\input{sections/6discussion}
\input{sections/acknowledgements}

% \clearpage
% \onecolumngrid
% \clearpage
\newpage
\bibliographystyle{unsrt}  
\bibliography{references}

\appendix
\onecolumngrid
\setcounter{figure}{0}
\renewcommand{\thefigure}{S\arabic{figure}}
\setcounter{equation}{0}
\renewcommand{\theequation}{S\arabic{equation}}
\input{sections/appendix}
\end{document}

%% file: preamble.tex
\usepackage[english]{babel}
\usepackage[utf8]{inputenc}
\usepackage[T1]{fontenc}
\usepackage[dvipsnames]{xcolor}
\usepackage[citecolor=MidnightBlue, colorlinks=true, linkcolor=MidnightBlue, urlcolor=MidnightBlue, linktocpage=true]{hyperref}
\usepackage{amsthm}
\usepackage{mathtools}
\usepackage{physics}
\usepackage{csquotes}
\usepackage{comment}
\usepackage[capitalise]{cleveref}
\usepackage[makeroom]{cancel}
\usepackage{qcircuit}
\usepackage[ruled,vlined]{algorithm2e}
\usepackage{bbm}
\usepackage{amsmath,bm}
\usepackage{enumitem}
\usepackage{booktabs}
\usepackage[caption=false]{subfig}
\usepackage{graphicx}
\usepackage{dblfloatfix}
\usepackage{mathtools}
\usepackage{float}
\usepackage{afterpage}

\DeclarePairedDelimiter\ceil{\lceil}{\rceil}

\newcommand{\id}{\text{I}}

\newcommand{\D}{\mathcal{D}}

\newcommand{\One}{\mathbbm{1}}

\newcommand{\Ub}{\mathbf{U}}

\newcommand{\rhob}{\boldsymbol{\rho}}
\newcommand{\thetab}{\boldsymbol{\theta}}

\newcommand{\St}{\widetilde{S}}

\DeclarePairedDelimiterX{\expectarg}[1]{[}{]}{%
  \ifnum\currentgrouptype=16 \else\begingroup\fi
  \activatebar#1
  \ifnum\currentgrouptype=16 \else\endgroup\fi
}

%% file: sections/1introduction.tex
\section{Introduction \label{sec:introduction}}

Noise is often considered to be one of the strongest adversaries of practical quantum computation. Decoherence effects due to a noisy environment can create errors in the final output of a circuit, destroying the advantage of many quantum algorithms. In contrast, noise is also what underlies stochastic processes, and is therefore a crucial element in classical computing, solving tasks such as sampling and optimization. In quantum systems, noise has also been shown to be a useful resource in several applications: carefully engineered dissipative processes can lead to universal quantum computation~\cite{verstraete2009quantum}, shot-noise in the measurement process can drive variational algorithms out of local minima~\cite{kubler2020adaptive, sweke2020stochastic}, and amplitude-damping channels can significantly improve quantum autoencoders for mixed states~\cite{cao2020noise}. We investigate in the present paper a novel way to exploit noise in near-term quantum devices, with the objective of studying a central task in quantum computing: thermal state preparation.

Placing $N$ qubits in an open quantum processing unit driven by a Hamiltonian $H \in \mathbb{C}^{2^N \times 2^N}$ and an effective temperature $T=\frac{1}{\beta}$, the system will asymptotically reach a thermal equilibrium state, given by the quantum Gibbs distribution
\begin{equation}
    \rho_\beta = \frac{1}{Z} e^{-\beta H} \label{eq:thermal_state}
\end{equation}
where $Z = \Tr[e^{-\beta H}]$ is the partition function.
Efficiently preparing a thermal state on a quantum computer is a problem of broad practical importance, with applications ranging from quantum chemistry and many-body physics simulations in an open environment~\cite{cao2019quantum, whitfield2013introduction, lee2020neural} to semi-definite programming~\cite{brandao2017quantum,van2017quantum} and quantum machine learning~\cite{kieferova2017tomography, amin2018quantum}.
However, sampling from a general Gibbs distribution is a computationally hard task for classical computers, due to the complexity of calculating the partition function~\cite{jerrum1993polynomial}. Most techniques rely on Monte-Carlo Markov Chain (MCMC) algorithms, which are often provably efficient only above a certain threshold temperature~\cite{eldan2020spectral}. 

Many algorithms have been proposed to prepare the thermal state on a quantum computer. A growing body of work has suggested using variational algorithms to solve the task of thermal state preparation on Noisy Intermediate Scale Quantum (NISQ) devices. Since a unitary circuit acting on the zero-state cannot directly output a mixed state, most variational thermalization methods consist either in preparing a purification of the thermal state and tracing out the ancillary qubits at the end of the circuit~\cite{hsieh-variational-tfd,nasa-thermal-state,ibm-imaginary-time, entropy-taylor}, or in choosing an appropriate mixed state as input~\cite{verdon2017quantum, product-spectrum-ansatz,verdon2019quantum}. 

One of the main challenges associated to those methods is to design an appropriate cost function to be minimized during the variational training loop. While the ground-state of a Hamiltonian can be prepared by minimizing the average energy of the state, the thermal state can be prepared by minimizing the \textit{free energy} $F=H-TS$ of the state, where $S=-\Tr[\rho\log(\rho)]$ is its Von Neumann entropy. However, the Von Neumann entropy is not an observable and can often only be computed approximately~\cite{entropy-taylor, wiebe-entropy-fourier}. A second problem is the need for additional qubits, which can be costly in near-term devices. Finally, none of those methods take into account the noise of the circuit, which can change the spectrum of the final state and affect the performance of the preparation algorithm~\cite{franca2020limitations}.

\begin{figure}[t]
    \subfloat[NAVQT ansatz.\label{fig:ansatz}]{
        \includegraphics[width=0.97\linewidth]{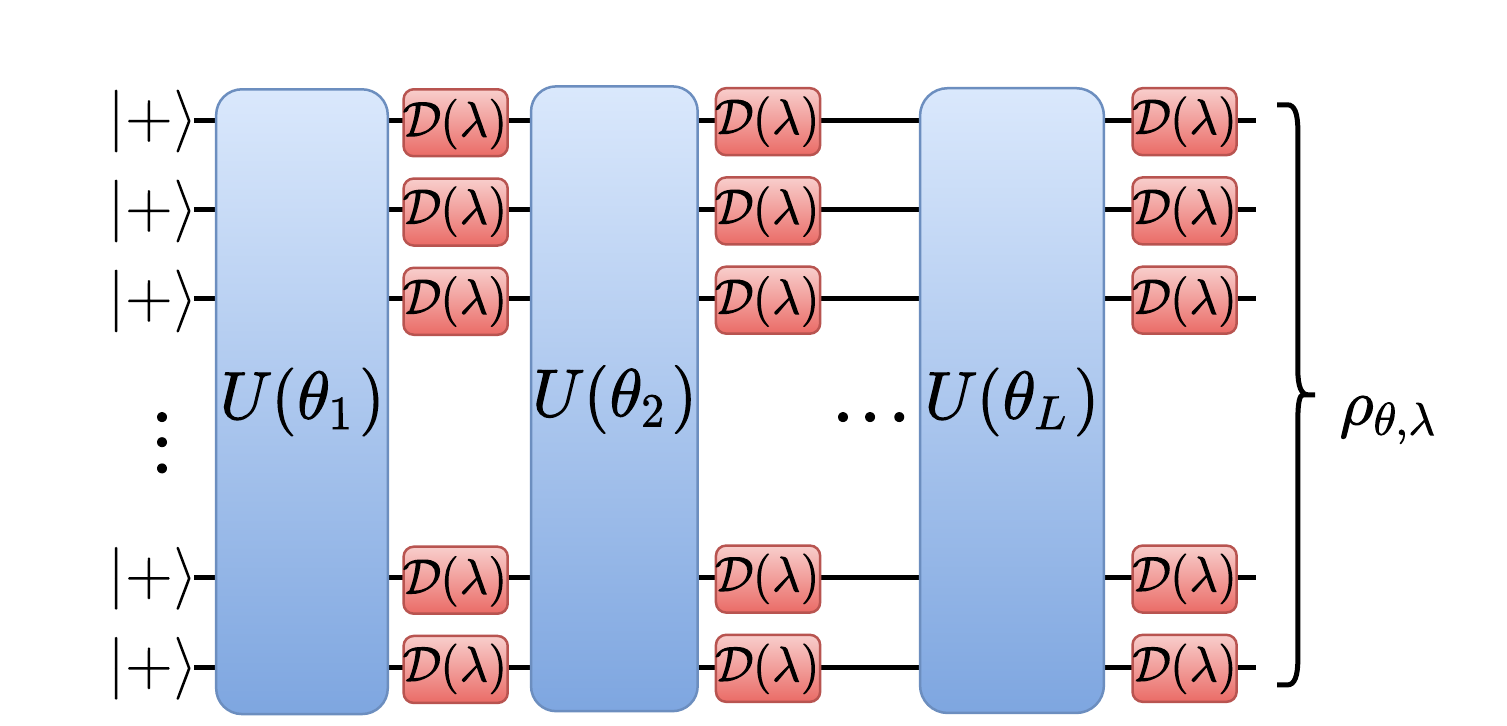}
    }\\
    \subfloat[Approximation for entropy calculations \label{fig:ansatz_approx}]{
        \includegraphics[width=0.97\linewidth]{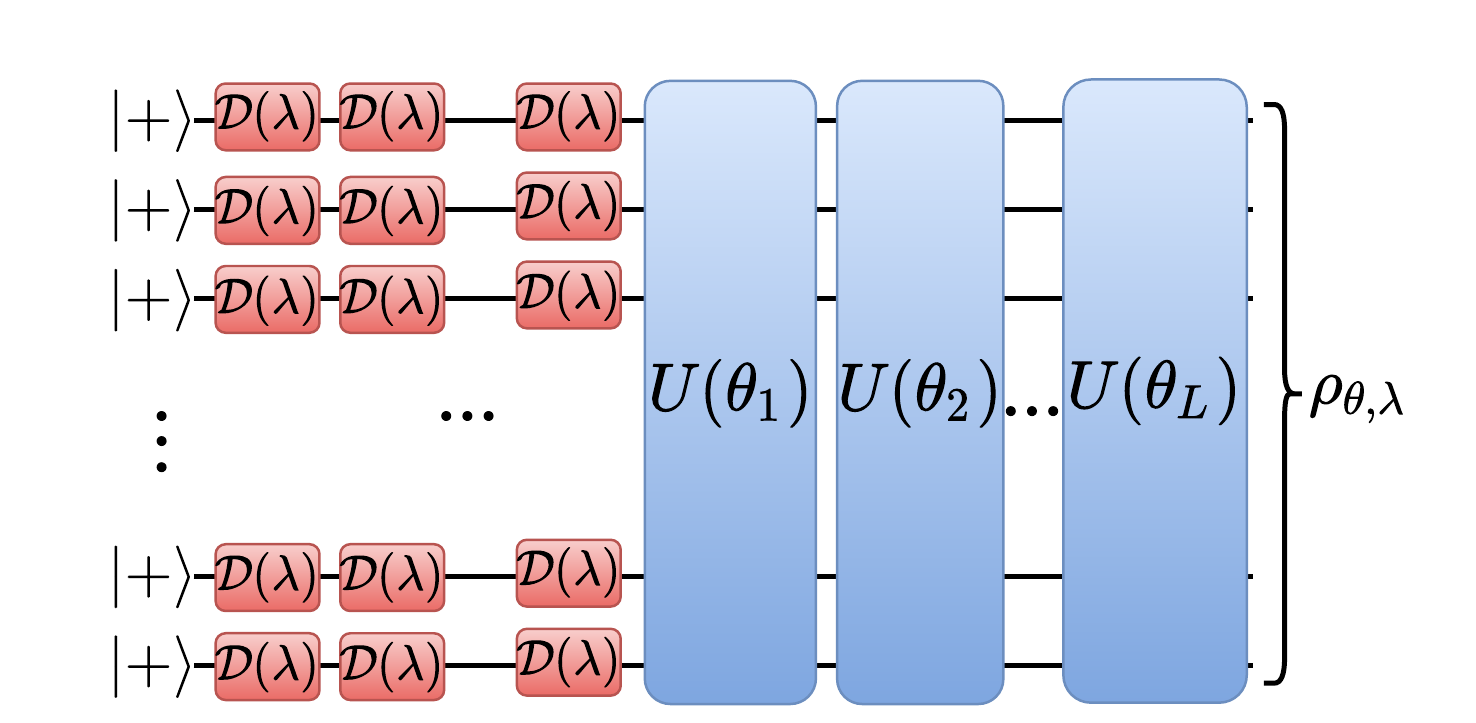}
    }
    \caption{Illustration of circuit components used in NAVQT. \textbf{(a)} General NAVQT ansatz: a sequence of unitary layers $U(\theta_i)$ followed by depolarizing gates $\mathcal{D}(\lambda)$ on each qubit. \textbf{(b)} Approximation used in the free-energy calculations}\label{fig:circuit-components}
\end{figure}

In this paper, we propose a new method that we call \textit{Noise-Assisted Variational Quantum Thermalizer} (NAVQT). Our algorithm assumes the ability to control the noise in the system down to some minimal noise level determined by the hardware. This type of control has been demonstrated in the context of error mitigation, where noise is increased in order to perform zero-noise extrapolation~\cite{temme2017error, endo2020hybrid}. More precisely, we construct a variational circuit with a parametrized depolarizing channel after each layer of unitary gates, as illustrated in \cref{fig:ansatz}. To simplify the optimization process, we have only considered the case where all the depolarizing parameters take the same value. By varying both gate and noise parameters, we seek to minimize the free energy of the final state.

In order to compute the free energy (and its gradient), we derive an analytical expression for the entropy of a slightly different circuit: one where all the depolarizing gates have been displaced at the beginning of the circuit, as shown in \cref{fig:ansatz_approx}. Using this approximation, we can compute the gradient of the free energy with respect to both the noise and the unitary parameters. While this might be a rough estimate of the actual gradient, we show that this approximate optimization problem exhibits similar performance as when minimizing the true free energy.

We then empirically investigate our algorithm on three different types of Hamiltonians: the Ising chain, with and without a transverse field, and the Heisenberg model. For each model, we consider both uniform coefficients and coefficients drawn from a standard normal distribution, and train our variational algorithm for several choices of hyperparameters (number of layers, learning rates, initialization, etc.). To study the performance of our approach, we extract the fidelity of the prepared state compared to the actual thermal state for a range of different temperatures. 

Our results reveal different patterns. On the one hand, fidelities above $0.9$ are reached for uniform Ising chains, with and without a transverse field, for all temperatures and system sizes up to $N=7$. On the other hand, the performance tend to decrease with the system size and for specific ranges of temperatures, with fidelities that can get below $0.7$ for some of the most complex systems tested in this work.

Our paper is organized as follows. We start by reviewing previous work on variational thermalization in \cref{sec:related}. We then introduce NAVQT in \cref{sec:navqt}. We follow this up by a description of our experiments in \cref{sec:method}, and present our results in \cref{sec:result}. Finally, we discuss our work and provide ideas for future studies in \cref{sec:discussion}.

%% file: sections/2relatedwork.tex
\section{Related Work \label{sec:related}} 

Variational circuits have recently been proposed for thermal state preparation, due to the existence of a natural cost function for this task: the free energy. 
Using variational circuits to prepare a thermal state presents two challenges specific to this task: 1) finding an ansatz that can prepare mixed states, 2) finding a scalable optimization strategy.

\subsection{Choice of the ansatz}

A common approach to VQT consists in preparing a purification of the thermal state using a variational circuit that acts on $2N$ qubits---$N$ system qubits and $N$ ancilla/environment qubits---, and tracing the ancilla qubits out at the end of the circuit~\cite{hsieh-variational-tfd,nasa-thermal-state,ibm-imaginary-time, entropy-taylor}. An example of purification often considered in the literature is the thermofield double (TFD) state \cite{hsieh-variational-tfd,nasa-thermal-state}. For a Hamiltonian $H$ and an inverse temperature $\beta$, it is given by
\begin{align} \label{eq:tfd}
    \ket{\text{TFD}}=\frac{1}{\sqrt{Z}} \sum_n e^{-\beta E_n / 2} \ket{n}_S \otimes \ket{n}_E
\end{align}
where the $\{E_n, \ket{n}\}_n$ are pairs of eigenvalue/eigenvector of $H$, and subscript $S$ and $E$ refers to the system and environment, respectively.
For instance, Refs.~\cite{hsieh-variational-tfd, nasa-thermal-state} use a Quantum Approximate Optimization Ansatz (QAOA) ansatz acting on $2N$ qubits to prepare the TFD state of the transverse-field Ising model, the XY chain, and free fermions. One advantage of this approach is the ability to simulate the TFD, which can be interesting in in its own right, for instance for studying black holes~\cite{thermofield-double}. The obvious disadvantage is that it requires twice as many qubits that the thermal state we want to simulate. A converse approach consists in starting with a mixed state $\rho_0$ and applying a unitary circuit ansatz on the $N$ qubits of the system. The initial $\rho_0$ can either be fixed~\cite{verdon2017quantum} or modified during the optimization process~\cite{product-spectrum-ansatz,verdon2019quantum}.
In Ref.~\cite{verdon2017quantum}, $\rho_0$ is the fixed thermal state of $H_I=\sum_{i=1}^N Z_i$, where $Z_i$ is the Pauli $Z$ operator applied to qubit $i$ of the system. It can easily be prepared using the purification
\begin{align}
    \bigotimes_j \sqrt{2 \cosh(\beta)} \sum_{b \in \{0,1\}^N} e^{(-1)^{1+b} \beta / 2} \ket{b}_S \ket{b}_{E}.
\end{align}
However, since the spectrum does not change when we apply the unitary ansatz, having a static $\rho_0$ freezes the spectrum of the final state. Therefore, if the spectrum of the thermal state we want to approximate is far from the spectrum of $\rho_0$, this approach will fail.
In Ref.~\cite{product-spectrum-ansatz}, they use the thermal state $\rho_0(\bm{\epsilon})$ of $H=\sum_{i=1}^n \epsilon_i P_i$, where $P_i=\frac{1-Z_i}{2}$ as an initial state and $\bm{\epsilon}=\{\epsilon_1,...,\epsilon_n \}$ are parameters optimized during the training process.
Finally, Ref.~\cite{verdon2019quantum} proposes to use a unitary with stochastic parameters to prepare $\rho_0$. More precisely,
\begin{align}
    \rho_0(\bm{\theta})=V(X_{\bm{\theta}})\ketbra{0} V(X_{\bm{\theta}})^\dag
\end{align}
where $V(\bm{x})$ is a unitary ansatz and $X_{\bm{\theta}} \sim p_{\bm{\theta}}$ is a random vector with parametrized density $p_{\bm{\theta}}$. The density $p_{\bm{\theta}}$ can be given by a classical model, such as an energy-based model (e.g. restricted Boltzmann machine) or a normalizing flow, which will be trained to get a $\rho_0$ with a spectrum close to the thermal state of interest. 

\subsection{Optimization strategies}

Once the ansatz has been fixed, the parameters within needs to be optimized. Two main approaches have been proposed in the literature: 1) explicitly minimizing the free energy, 2) using imaginary-time evolution. In the following, we describe both these methods.

\paragraph{Free energy methods}
The thermal state is the density matrix that minimizes the free energy. Therefore, in the same way as VQE uses the energy as a cost function, any thermal state preparation method can use the free energy as its cost function~\cite{hsieh-variational-tfd,nasa-thermal-state,verdon2017quantum,verdon2019quantum}. However, one main difference with VQE is that the free energy cannot be easily estimated. Indeed, the Von Neumann entropy term, as a non-linear function of $\rho$, cannot be turned into an observable, and doing a full quantum state tomography would be very costly. Several methods have been proposed to solve this challenge:
\begin{itemize}
    \item Computing several Renyi entropies $S_{\alpha}=\frac{1}{1-\alpha} \Tr\left[ \rho^{\alpha} \right]$ (using multiple copies of $\rho$) and approximating the Von Neumann entropy with them~\cite{hsieh-variational-tfd, d2021alternative}. 
    \item Computing the Von Neumann entropies locally on a small subsystem~\cite{hsieh-variational-tfd}
    \item Approximate the Von Neumann entropy by truncating its Taylor~\cite{entropy-taylor} or Fourier~\cite{wiebe-entropy-fourier} decomposition.
\end{itemize}
In our work, the entropy term does not come from a purification procedure, but from the presence of depolarizing gates in the circuits. This led us to propose a different type of approximation that we will study in \cref{sec:navqt}.

\paragraph{Imaginary-time evolution} Thermal state preparation can be seen as the application of imaginary-time evolution during a time $\Delta t=i\beta/2$ on the maximally-mixed state $\rho_m=\frac{1}{d} \id$, using the decomposition
\begin{align*}
    \rho_{\beta}=\left( \frac{1}{C} e^{-\beta H/2} \right) \left( \frac{1}{d} \id \right) \left(\frac{1}{C}e^{-\beta H/2}\right)
\end{align*}
This imaginary-time evolution can be simulated using a variational circuit and a specific update rule~\cite{imaginary-time-evolution-benjamin}. 
In Ref.~\cite{ibm-imaginary-time}, the authors use a variational circuit $U(\thetab)$ on $2N$ qubits, initialized such that
\begin{align*}
    U(\thetab_0) \ket{0}^{\otimes 2N} \approx \ket{\Phi^+}
\end{align*}
where $\Phi^+$ is a maximally-entangled state. An imaginary-time update rule with a small learning rate $\tau$ will lead to a unitary $U(\thetab_0)$ such that:
\begin{align*}
    U(\thetab_1) \ket{0}^{\otimes 2N} \approx \frac{1}{C} e^{-\tau H} \ket{\Phi^+}
\end{align*}
Repeating it during $k=\frac{\beta}{2}$ steps will give the state
\begin{align*}
    U(\thetab_k) \ket{0}^{\otimes 2N} \approx \frac{1}{C} e^{-\beta H / 2} \ket{\Phi^+}
\end{align*}
which will be the thermal state after tracing out the environment.
In Ref.~\cite{japanese-imaginary-time}, the authors also use imaginary-time evolution to prepare the thermal state, but manage to reduce the number of qubits to $N$ when the Hamiltonian is diagonal in the $Z$-basis. Finally, an ansatz-independent imaginary-time evolution method has been proposed for thermal state preparation~\cite{sun2021quantum, motta2020determining}. 

In this work, we optimize the ansatz parameters using the free energy approach. Adapting imaginary-time evolution to a noisy ansatz could however be an interesting alternative, that we let for future work.

%% file: sections/3navqt.tex
\section{Noise-assisted Variational Quantum Thermalization \label{sec:navqt}}

We introduce here the \textit{Noise-Assisted Variational Quantum Thermalizer} (NAVQT), a variational algorithm where depolarizing noise is used as the source of entropy for preparing the thermal state. We consider a noise model where each layer of unitary gates is followed by a one-qubit depolarizing channel
\begin{align}
    \D(\lambda)(\rho)=(1-\lambda) \rho + \lambda \frac{\id}{2},
\end{align}
where $\id$ is the identity matrix. The channel is represented in \cref{fig:ansatz}. 
For the purpose of this work, we consider that we have the same noise value $\lambda \in [\lambda_{\min},1]$ everywhere in the circuit, where $\lambda_{\min}$ is the minimum noise reachable by the hardware. 
We note $\rho_{\thetab,\lambda}$ the output of the noisy circuit with unitary parameters $\thetab$ and noise parameter $\lambda$, and want to find the optimal parameters $\{\thetab^*,\lambda^*\}$ such that $\rho_{\thetab^*,\lambda^*} \approx \rho_{\beta}$ where the latter is given by \cref{eq:thermal_state}. 

The thermal state $\rho_{\beta}$ can be approximated by minimizing the free energy of the system, given by:
\begin{align}
    F(\thetab,\lambda) = E(\bm{\theta},\bm{\lambda}) - \frac{1}{\beta} S(\thetab,\lambda) \label{eq:free_energy}
\end{align}
where 
\begin{equation}
    E(\thetab,\lambda) =\Tr[H \rho_{\thetab,\lambda}] \label{eq:energy}
\end{equation}
is the energy and 
\begin{equation}
    S(\thetab,\lambda)=-\Tr[ \rho_{\thetab,\lambda} \log(\rho_{\thetab,\lambda})] \label{eq:entropy}
\end{equation}
is the Von Neumann entropy of the state.

The energy term and its gradient are easy to evaluate: we can use the parameter shift-rule~\cite{schuld2019evaluating} to compute $\nabla_{\thetab} E(\thetab,\lambda)$, and the finite-difference method to calculate $\partial_{\lambda} E(\thetab,\lambda)$. The entropy term is much harder to evaluate as it is a non-linear function of the state. To approximate it, we consider the circuit where all the noise has been put at the beginning, as shown in \cref{fig:ansatz_approx}. While the resulting free energy will not be equal to the free energy of our original circuit in general, they tend to follow similar trajectories when varying the noise level (see Supplementary \cref{fig:entropy_energy_simulations}). The new entropy does not depend on $\thetab$ and can be calculated analytically as if there were no unitary gates. For a circuit with $N$ qubits and $m$ layers, this approximate entropy $\St(\lambda)$ is given by
\begin{align}
    \begin{split}
     \St(\lambda) &= - N \left((1-\lambda)^{m} + \frac{(1-(1-\lambda)^{m})}{d} \right) \cdot \\ 
     & \ln ((1-\lambda)^{m} + \frac{(1-(1-\lambda)^{m})}{d}) \\
        &+  \frac{(d-1)(1-(1-\lambda)^{m})}{d} \ln(\frac{(1-(1-\lambda)^{m})}{d})
    \end{split}
\end{align}
where $d=2^N$. The full derivation is given in the Supplementary material. Using this approximation, we get the following gradient-based update rule at each optimization step:
\begin{align}
    \thetab^{(n+1)} &= \thetab^{(n)} - \eta_{\theta} \nabla_{\thetab} E(\thetab,\lambda) \\
    \lambda^{(n+1)} &= \lambda^{(n)} - \eta_{\lambda} \left(\nabla_\lambda E(\thetab,\lambda) - \frac{1}{\beta} \nabla_\lambda \St(\lambda)\right)
\end{align}
where $\eta_{\theta}$ and $\eta_{\lambda}$ are the learning rates for $\thetab$ and $\lambda$, respectively.

%% file: sections/4method.tex
\section{Methods}
\label{sec:method}
%The methods should be succinct and must not contain subheadings. Topical subheadings are allowed. Authors must ensure that their Methods section includes adequate experimental and characterization data necessary for others in the field to reproduce their work.
In this section, we describe the method used to evaluate NAVQT. All quantum circuit simulations are done in Cirq~\cite{cirq} and TensorFlow-Quantum~\cite{broughton2020tensorflow}.

\subsection{Ansatz}
\label{sec:ansatz}

\begin{figure}[t]
    \includegraphics[width=0.25\textwidth]{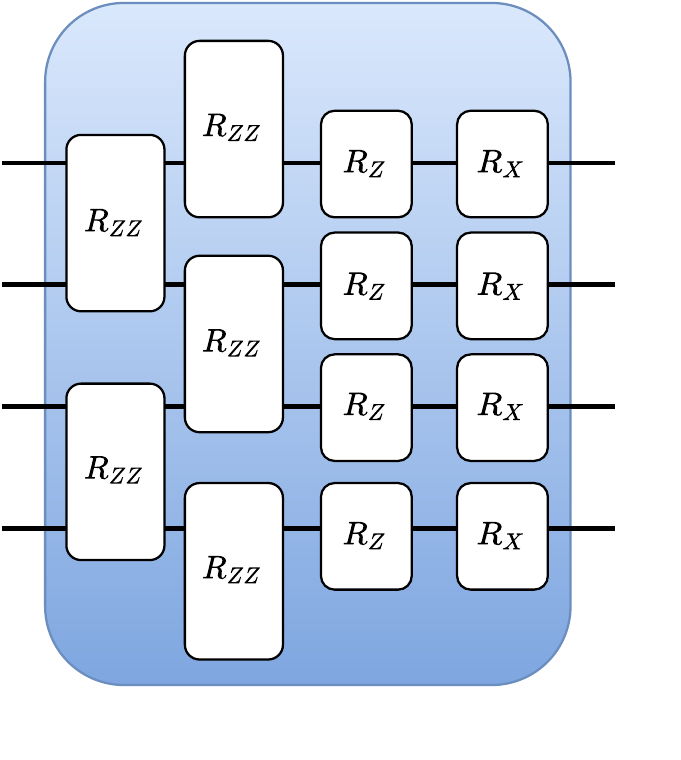}
    \caption{A layer of the unitary ansatz used in our experiments (for $N=4$ qubits), inspired by QAOA for the 1D Ising model. $R_Z$ and $R_X$ are parametrized rotations around the corresponding axis, and $R_{ZZ}=e^{-i\theta Z_iZ_j}$.
    \label{fig:unitary}}
\end{figure}

For the unitary layers of our circuit, we chose an ansatz inspired by the Quantum Approximate Optimization Ansatz (QAOA) applied to the Ising chain Hamiltonian~\cite{farhi2014quantum}. More precisely, if we define a problem Hamiltonian
\begin{equation}
    H_P = - \sum_i Z_i Z_{i+1} -  \sum_i Z_i
\end{equation}
and a mixing Hamiltonian
\begin{equation}
    H_M = - \sum_i X_i,
\end{equation}
the QAOA ansatz with $p$ layers is given by
\begin{equation}
    U(\vb{\gamma}, \vb{\beta})= e^{i\beta_p H_M} e^{i\gamma_p H_P} \dotsc e^{i\beta_1 H_M} e^{i\gamma_1 H_P}
    \label{eq:qaoa}
\end{equation}
This ansatz, whose explicit construction is represented in \cref{fig:unitary}, has been well-studied in the context of ground-state preparation~\cite{wierichs2020avoiding} and has been shown to be universal~\cite{morales2020universality} in the limit $p \rightarrow \infty$.
We test two different versions of this ansatz. In the first one, denoted \textit{restricted QAOA}, gates of the same type from a given layer share the same parameters $\beta_i$ and $\gamma_i$. In the second version, which we call \textit{flexible QAOA}, every gate has its own parameter.

We ran some preliminary tests to verify that this unitary ansatz is at least able to express the ground-state of all the systems tested in our work, and found it to be the case when the number of layers is fixed at $\ceil{\frac{N}{2}}$. Hence the noisy ansatz should in principle be able to represent the correct thermal state for large $\beta$, by setting $\lambda=0$ and fitting the unitary parameters corresponding to the ground-state. Moreover, NAVQT is also able to represent the maximally-mixed state, corresponding to a low $\beta$, by setting $\lambda=1$. The ability of the ansatz to learn intermediate temperatures is an open question, that we tackle in our numerical analysis.

\subsection{Hyperparameters}

Since the choice of hyperparameters can have a substantial impact on the performance of variational circuits~\cite{wierichs2020avoiding}, we perform a grid-search to reduce the potential negative effects resulting from a single design choice. Hence we try all combinations in the search space defined by
\begin{itemize}
    \item Restricted QAOA and flexible QAOA
    \item Initial noise level: $\lambda=\{10^{-8},0.001,0.1\}$ 
    \item Unitary learning rate: $\eta_{\theta}=\{0.01,0.4\}$ 
    \item Noise learning rate: $\eta_{\lambda}=\{0.0001,0.1\}$ 
    \item Seeds for unitary parameters: $[0;4]$ 
\end{itemize}
We run our algorithm for $N\in[3;7]$ qubits and for maximum $1000$ iterations. To test the performance across temperatures, we take 10 different betas in the interval $\beta \in [10^{-3};10^2]$, namely $
\{0.001,0.1,0.25,0.5,0.75,1.0,2.0,5.0,10.0,100.0\}$. We initialize the unitary parameters by sampling from a uniform distribution in the interval $[0.0001, 0.05]$ as done in~\cite{wierichs2020avoiding}. Finally, we extract the solution that gives the lowest (approximated) free energy among all the tested hyperparameters and initializations. We also include the same grid-search using finite-difference on the true free-energy in Supplementary \cref{fig:true-free-energy-optimization}.

\subsection{Noisy circuit simulation}

To simulate the noise in our circuit, we use the fact that depolarizing gates can also be written as~\cite{preskill1998lecture}
\begin{equation}
    \D(\lambda)(\rho) = \left(1-\frac{3 \lambda}{4}\right) \rho + \frac{\lambda}{4} \left(X \rho X + Y \rho Y + Z \rho Z \right)
\end{equation}
which can be interpreted as applying a random Pauli error with probability $p=\frac{3\lambda}{4}$ and nothing with probability $p=1-\frac{3\lambda}{4}$.
We can therefore simulate depolarizing gates as stochastic mixtures over unitary circuits containing errors. More precisely, if we sample $K$ unitaries $\Ub^{(k)}$, each being a combination of the unitary part of the ansatz and some random errors, we can extract the corresponding density matrix as:
\begin{equation}
    \rhob_{\text{out}} 
    \approx 
    \frac{1}{K}
    \sum_{k=1}^K
    \Ub^{(k)}
    \rhob_{in}
    \left(\Ub^{(k)}\right)^\dagger
\end{equation}
We found that taking a sample size of $K= 500 N$ was sufficient to get stable gradients and reach the maximum entropy $S \leq \log 2^N$. However, we also found that $K$ could be smaller, especially when $\beta$ was large and hence the target entropy was low.

\subsection{Performance metric}

For each experiment, we report the fidelity
\begin{equation}
    F(\rho_1,\rho_2) = \Tr[\sqrt{\sqrt{\rho_1}\rho_2\sqrt{\rho_1}}]
\end{equation}
between the thermal state and the output state of the trained circuit. Tracking the fidelity requires us to compute the true thermal state $\rho_{\beta}$ for each Hamiltonian $H$ and temperature $\beta$. In practice, taking the exponential of a matrix containing potentially large numerical values (e.g. when $\beta$ is large) can result in numerical issues. To alleviate those issues, we calculate the thermal state density matrix $\rho_\beta$ by taking the log on both sides of \cref{eq:thermal_state} and using the log-sum-exp trick~\cite{blanchard2021accurately}:

\begin{align}
\log \rho_\beta &= \log e^{-\beta H} - \log  \Tr[e^{-\beta H}] \\
&= -\beta H - \log   \sum_i e^{-\beta \lambda_i}  \\
&= -\beta H - \left(-\beta c + \log   \sum_i e^{-\beta (\lambda_i - c)} \right) 
\end{align}
where $c$ is the largest eigenvalue of $H$.

\subsection{Models}

We evaluated our algorithm on three different models: the Ising chain, with and without a transverse field, and the Heisenberg model. 
For each model, we considered two cases: when the coefficients $J_i = h_i = 1$ for all $i$, denoted the \textit{uniform} version, and when $J_i, h_i \sim \mathcal{N}(0,1)$ for all $i$, denoted the \textit{random} version. Between five seeds for the random version, we pick the Hamiltonian with the lowest spectral gap as this could considered the hardest Hamiltonian.

\paragraph{Ising chain}
The 1D Ising model, or Ising chain (IC), considers a set of spins on a chain such that all spins have exactly two coupled neighbors when considering $N > 2$. The Hamiltonian associated with such system is given by
\begin{equation}
    H_{\text{IC}} = - \sum_i J_{i} Z_i Z_{i+1} -  \sum_i h_{i} Z_i \label{eq:ising-eq}
\end{equation}
where $Z_i$ is the Pauli $Z$ operator acting on qubit $i$. 

\paragraph{Transverse field Ising chain}
The transverse-field Ising chain (TFI) adds quantum effects to the previous model by including some non-diagonal terms in its Hamiltonian. It is defined as
\begin{equation}
    H_{\text{TFI}} = - \sum_i J^{Z}_{i} Z_i Z_{i+1} - \sum_i h^{Z}_{i} Z_i - \sum_i h^{X}_{i} X_i
\end{equation}
where $X_i$ is the Pauli $X$ operator acting on qubit $i$.

\paragraph{Heisenberg model}
Finally, we consider the 1D Heisenberg model, whose Hamiltonian is given by
\begin{equation}
\begin{split}
    H_{\text{Heisenberg}}& = - \sum_i J^{Z}_{i} Z_i Z_{i+1} - \sum_i J^{X}_{i} X_i X_{i+1} \\
    &- \sum_i J^{Y}_{i} Y_i Y_{i+1} - \sum_i h^{X}_{i} X_i
\end{split}
\end{equation}
The Heisenberg model is of fundamental importance in the study of quantum materials~\cite{billoni2005spin, gong2014emergent, jepsen2020spin, rodriguez2021turbulent} and is therefore a standard benchmark for thermal state preparation methods~\cite{powers2021exploring, sun2021quantum, motta2020determining}.

%% file: sections/5results.tex
\section{Results \label{sec:result}}
\begin{figure}[t!]
    \subfloat[$\beta=0.1$]{
         \includegraphics[width=0.96\linewidth]{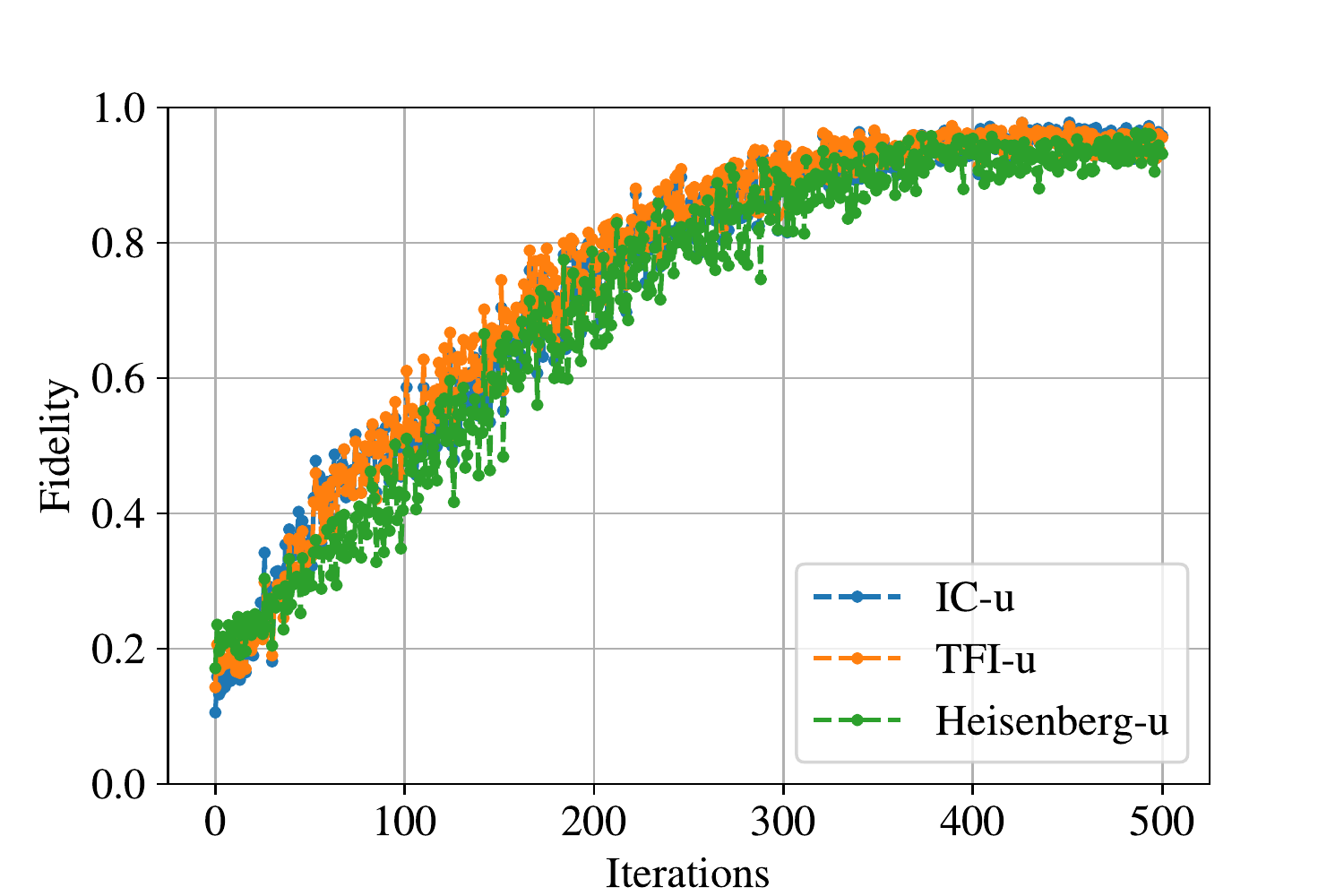}
    }\\
    \subfloat[$\beta=0.5$]{
         \includegraphics[width=0.96\linewidth]{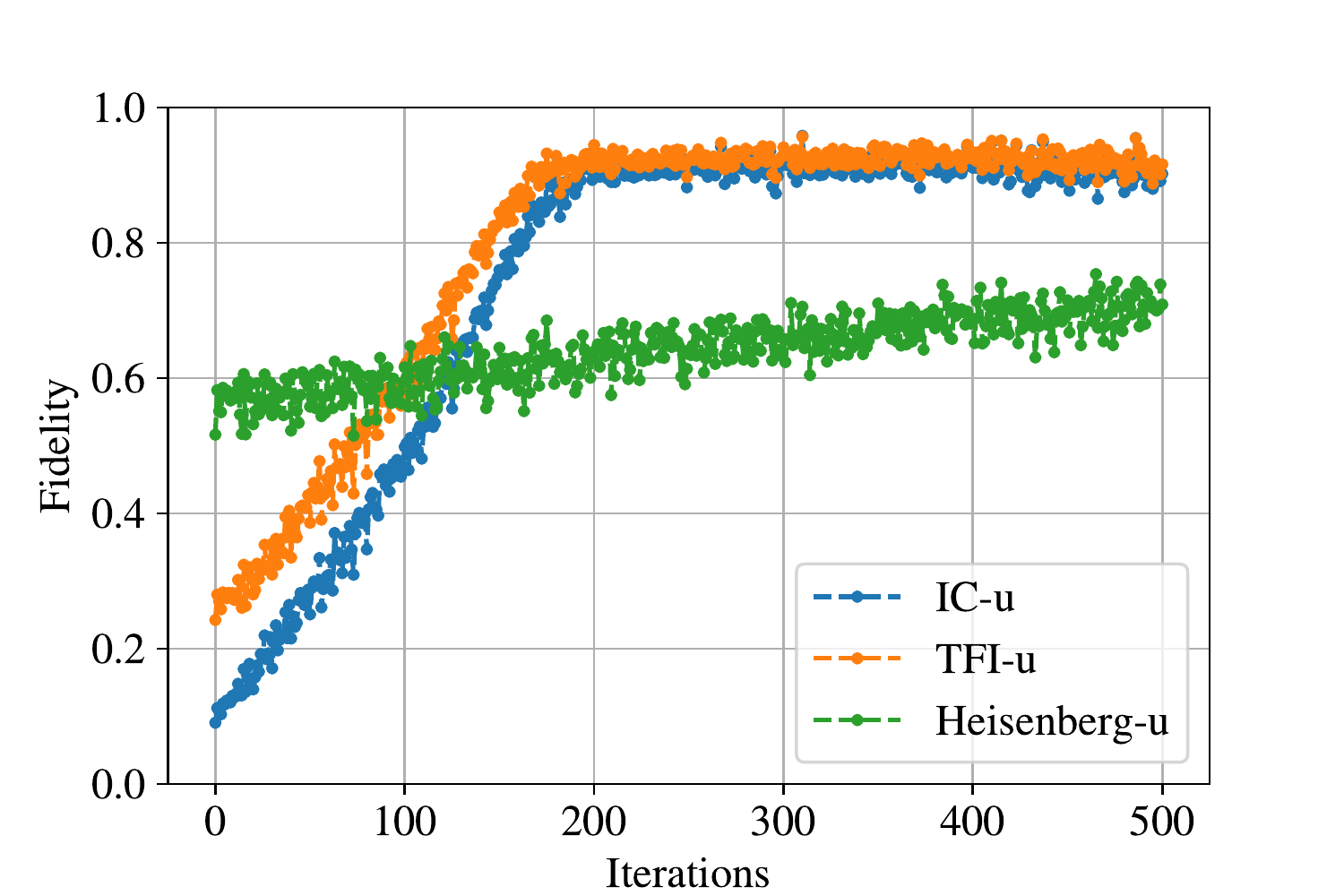}
    }\\
    \subfloat[$\beta=10$]{
         \includegraphics[width=0.96\linewidth]{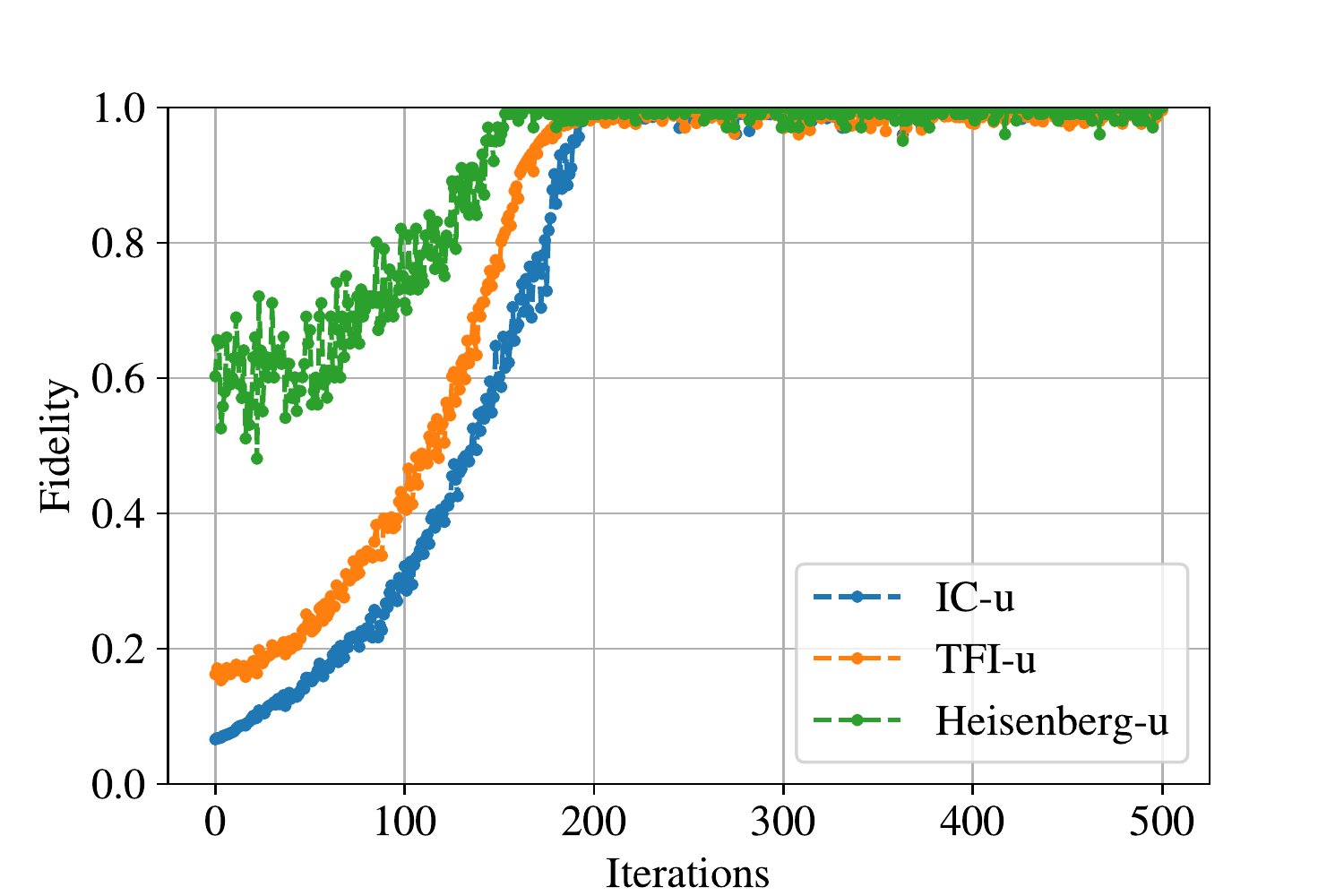}
    }
    \caption{Optimization curves for the three models with uniform coefficients and $N=4$. We observe in all the cases a constant increase of the fidelity, showing that minimizing the approximate free-energy cost function tends to result in a maximization of the fidelity. It also shows that the final result found by the algorithm is always significantly better than the random initialization.}
    \label{fig:learning-curves}
\end{figure}

\begin{figure*}
    \subfloat[Classical Ising chain with uniform coefficients. \label{fig:beta-vs-F-ising-u}]{
        \includegraphics[width=0.48\linewidth]{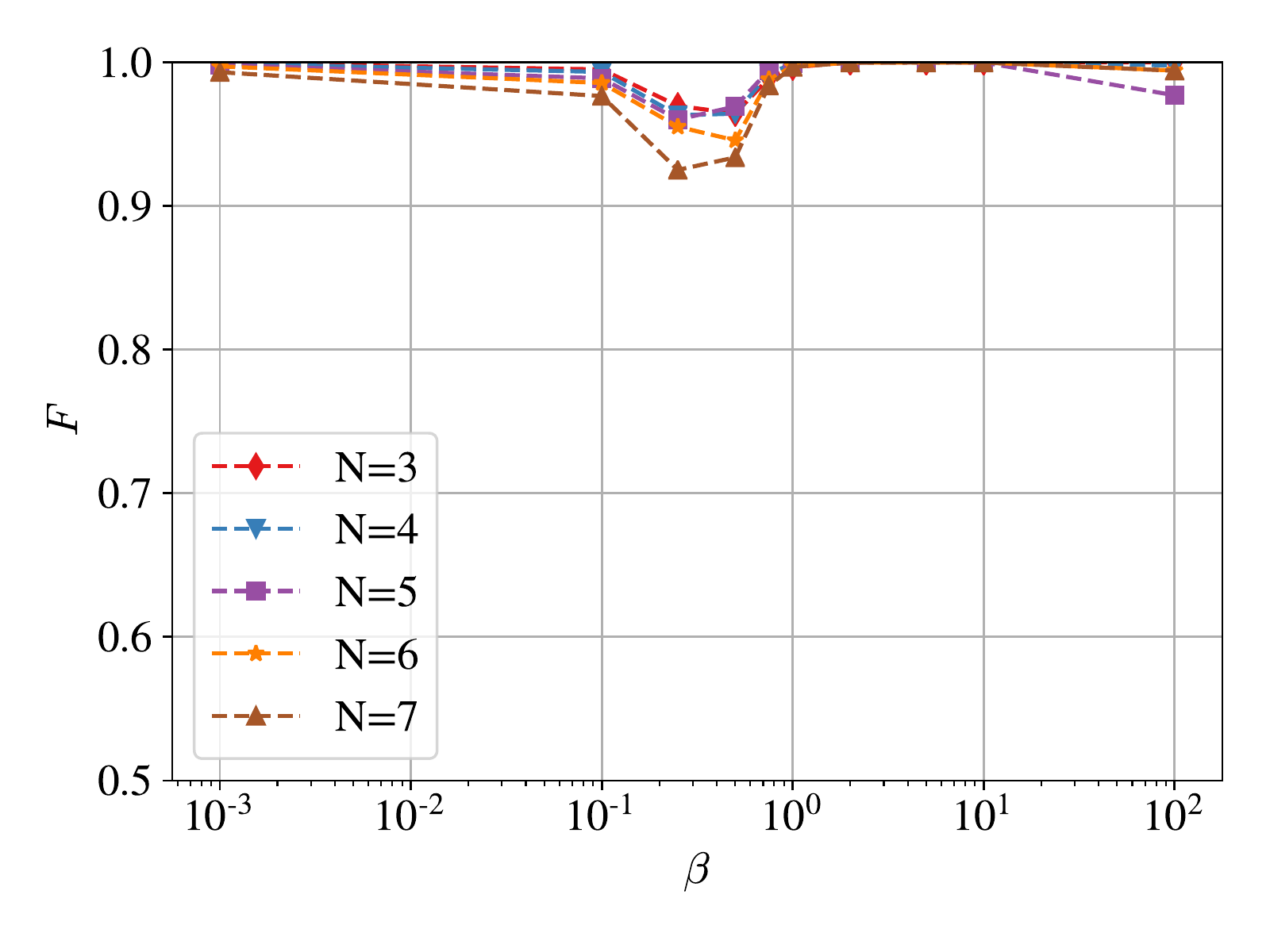}
    }
    \subfloat[Classical Ising chain with random coefficients.\label{fig:beta-vs-F-ising-r}]{
        \includegraphics[width=0.48\linewidth]{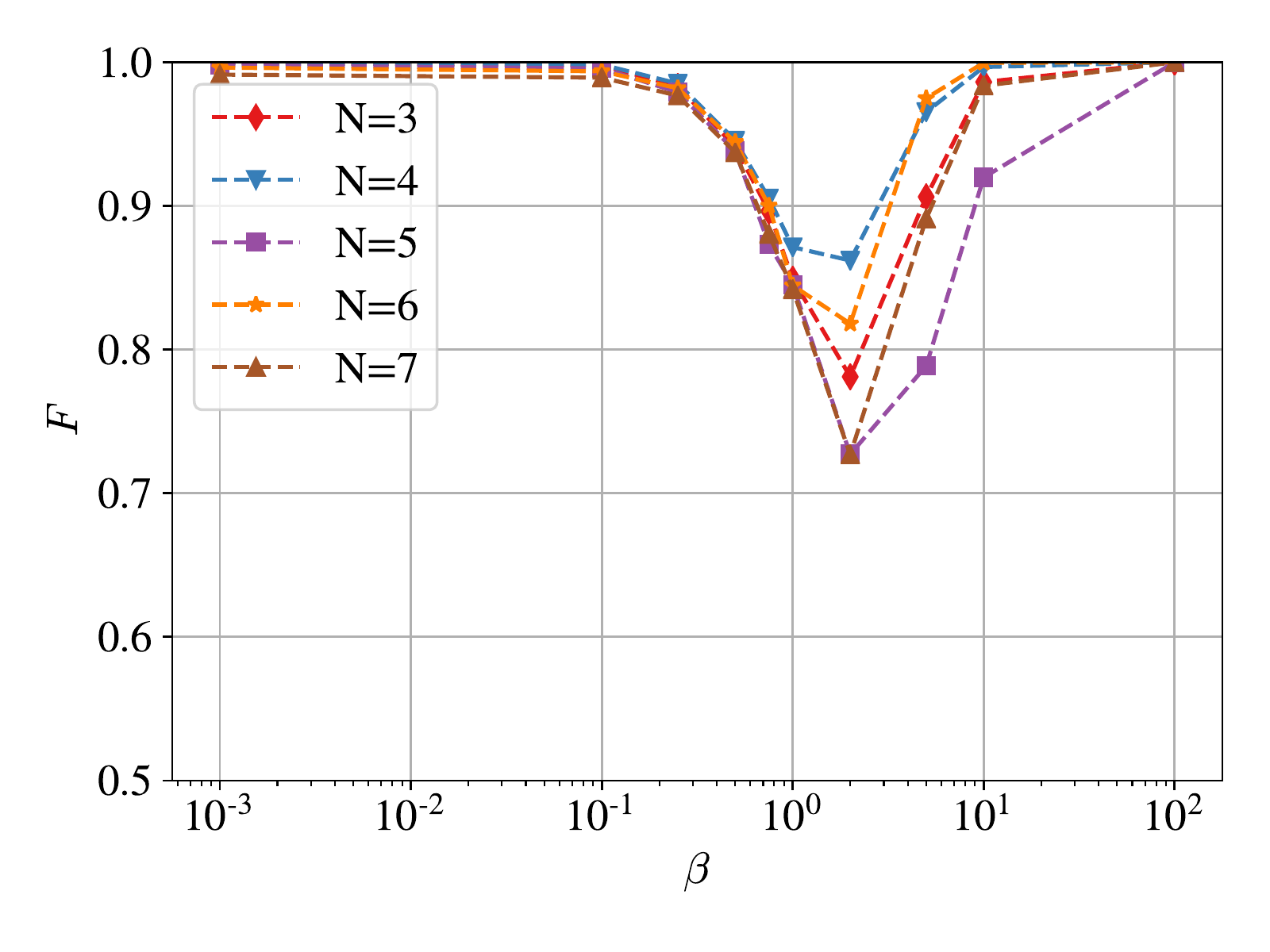}
    } \\
    \subfloat[Transverse-field Ising chain with uniform coefficients.\label{fig:beta-vs-F-tfi-u}]{
        \includegraphics[width=0.48\linewidth]{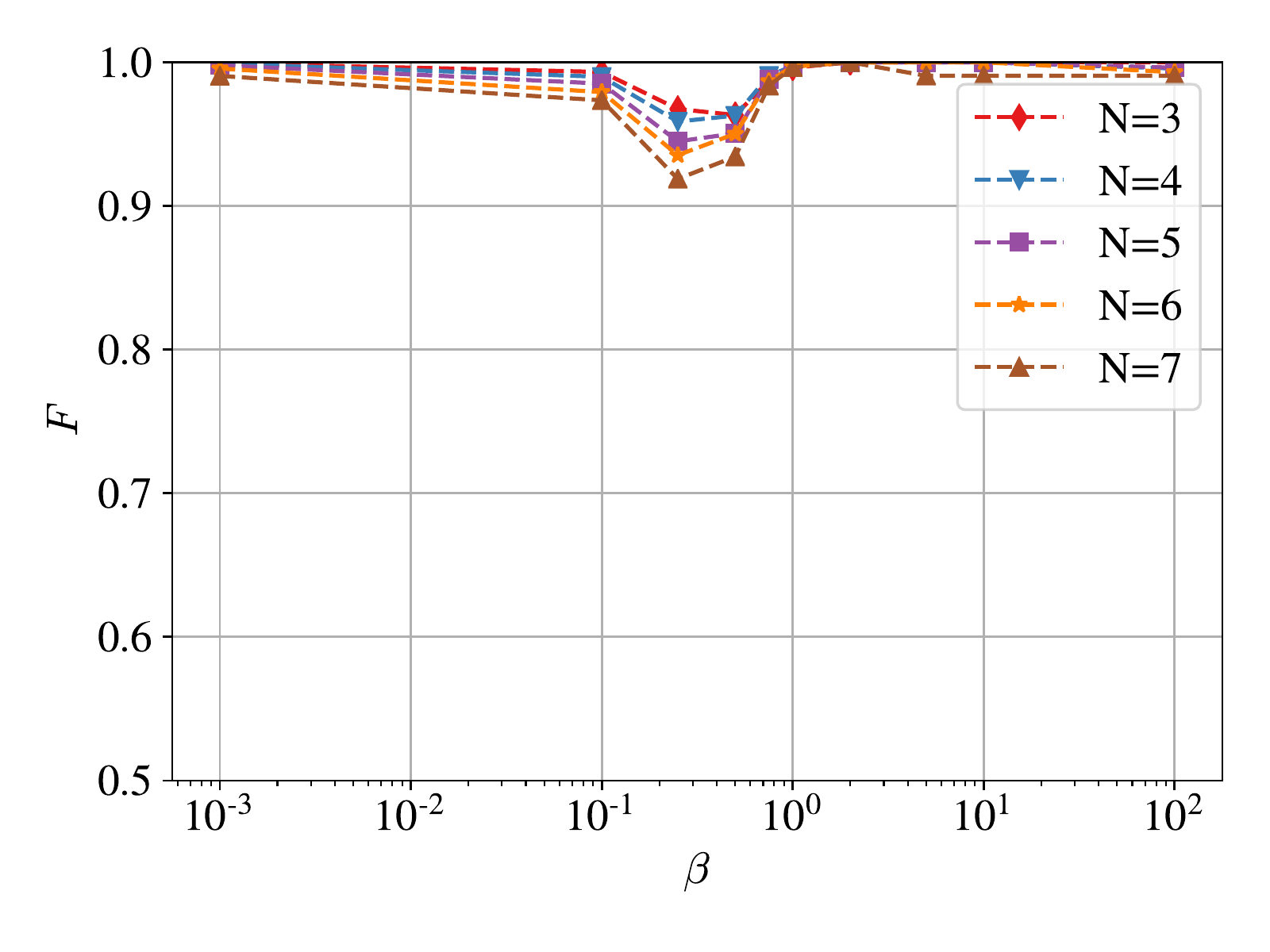}
    }
    \subfloat[Transverse-field Ising chain with random coefficients.\label{fig:beta-vs-F-tfi-r}]{
        \includegraphics[width=0.48\linewidth]{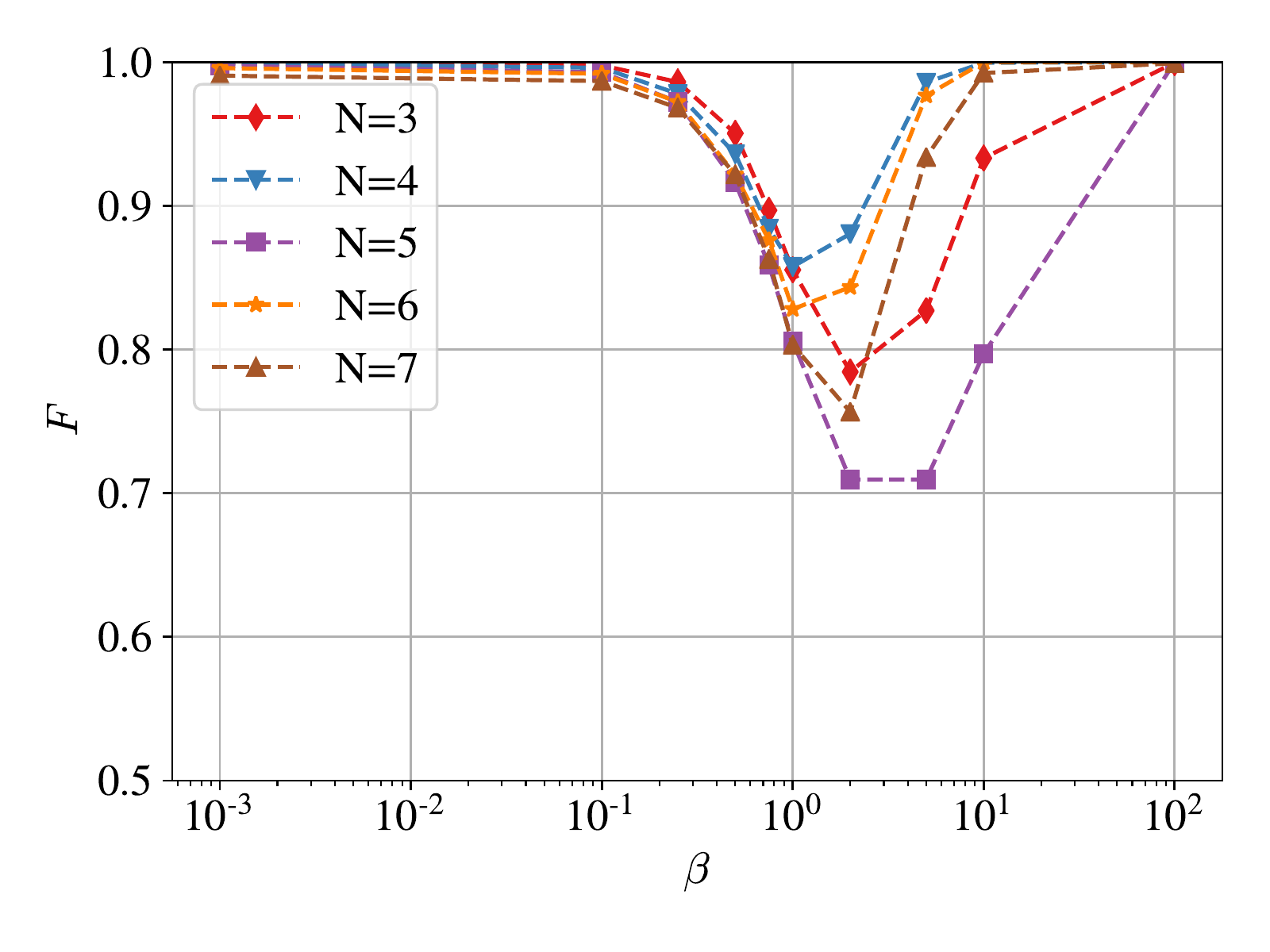}
    } \\
    \subfloat[1D Heisenberg model with uniform coefficients.\label{fig:beta-vs-F-heisen-u}]{
        \includegraphics[width=0.48\linewidth]{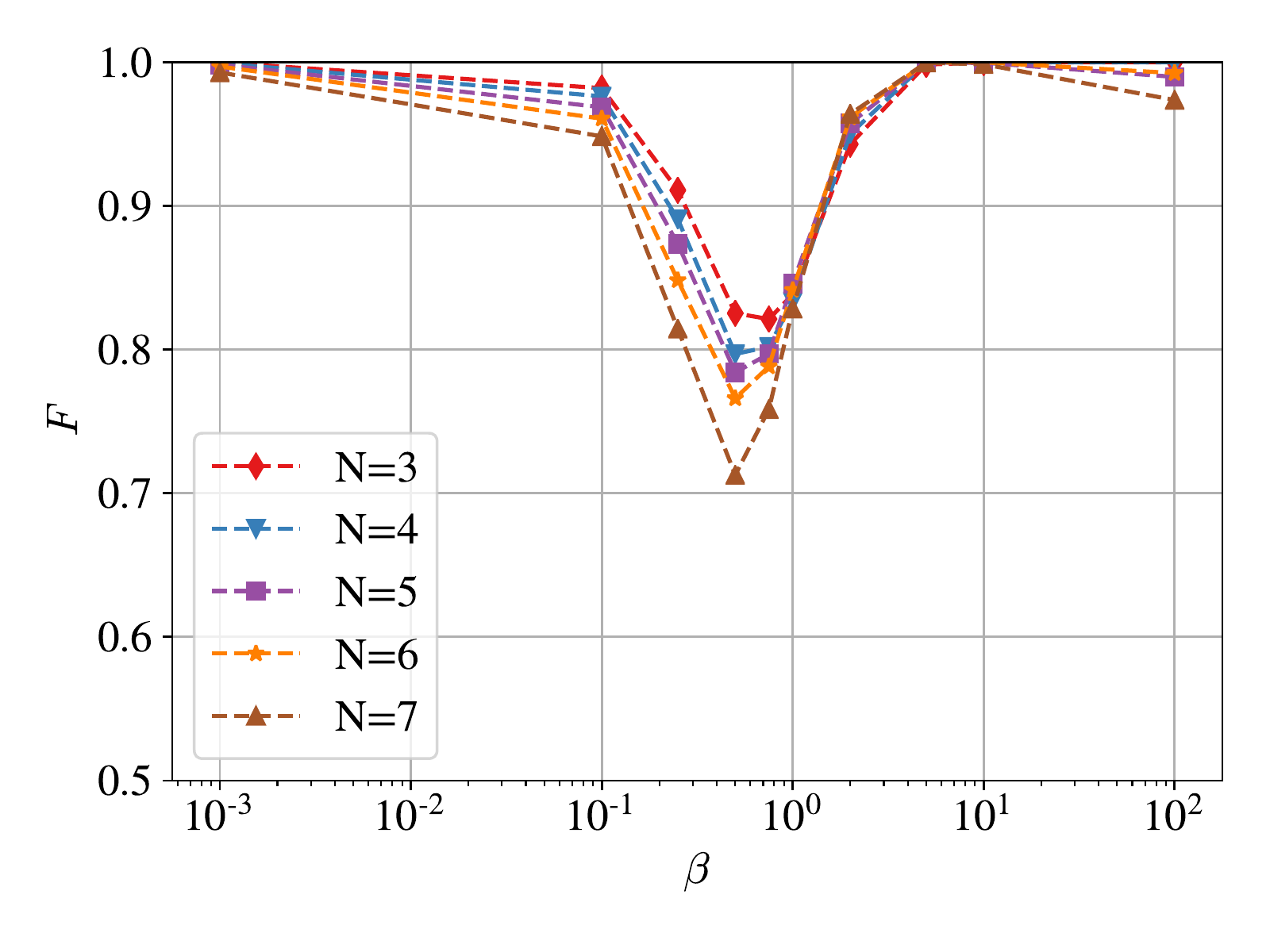}
    }
    \subfloat[1D Heisenberg model with random coefficients.\label{fig:beta-vs-F-heisen-r}]{
        \includegraphics[width=0.48\linewidth]{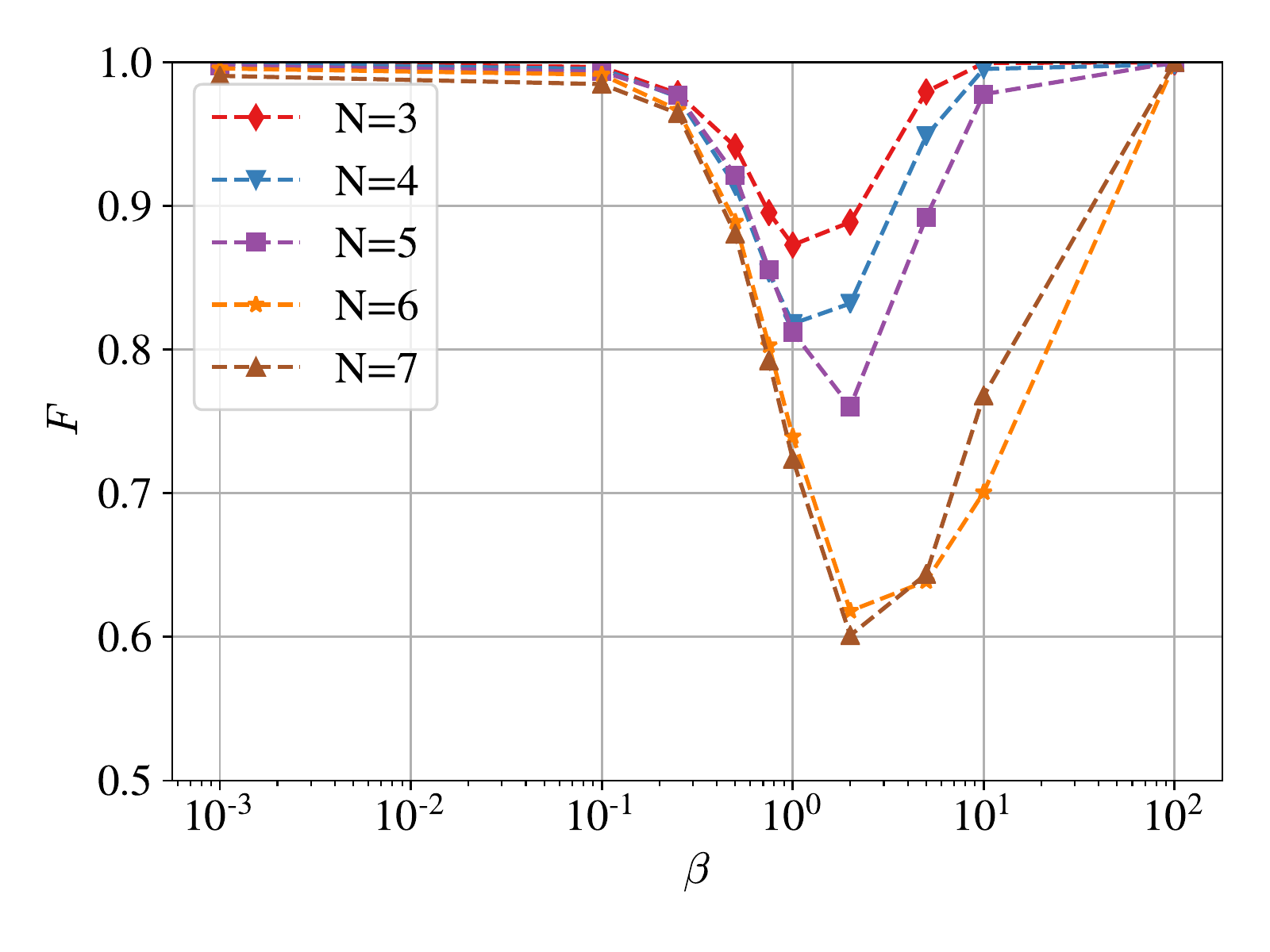}
    }

    \caption{Fidelities obtained using NAVQT as a function of the inverse temperature $\beta$, for various models and system sizes. For all the models, we observe that the algorithms reaches a high fidelity for low and high temperature, while it tends to decrease at intermediate temperatures. Overall, good performance is obtained at all temperatures for the two types of uniform Ising chains, while lower fidelities are reached with the other models.} \label{fig:beta-vs-F}
\end{figure*}

\begin{figure*}
    \centering
    \subfloat[Classical Ising chain with uniform coefficients.\label{fig:beta-vs-lambda-ising-u}]{
        \includegraphics[width=0.49\linewidth]{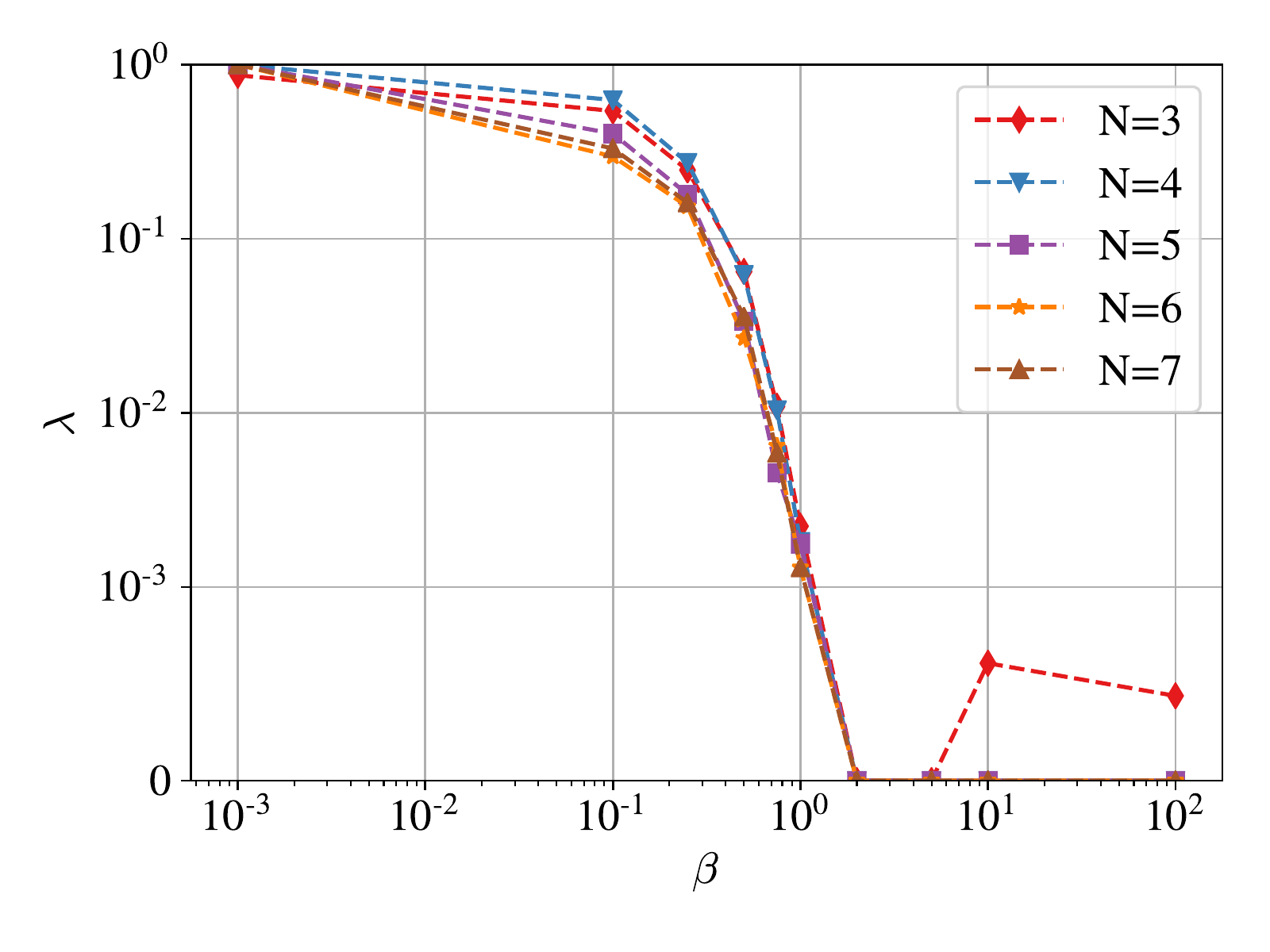}
    }
    \subfloat[Classical Ising chain with random coefficients.\label{fig:beta-vs-lambda-ising-r}]{    
        \includegraphics[width=0.49\linewidth]{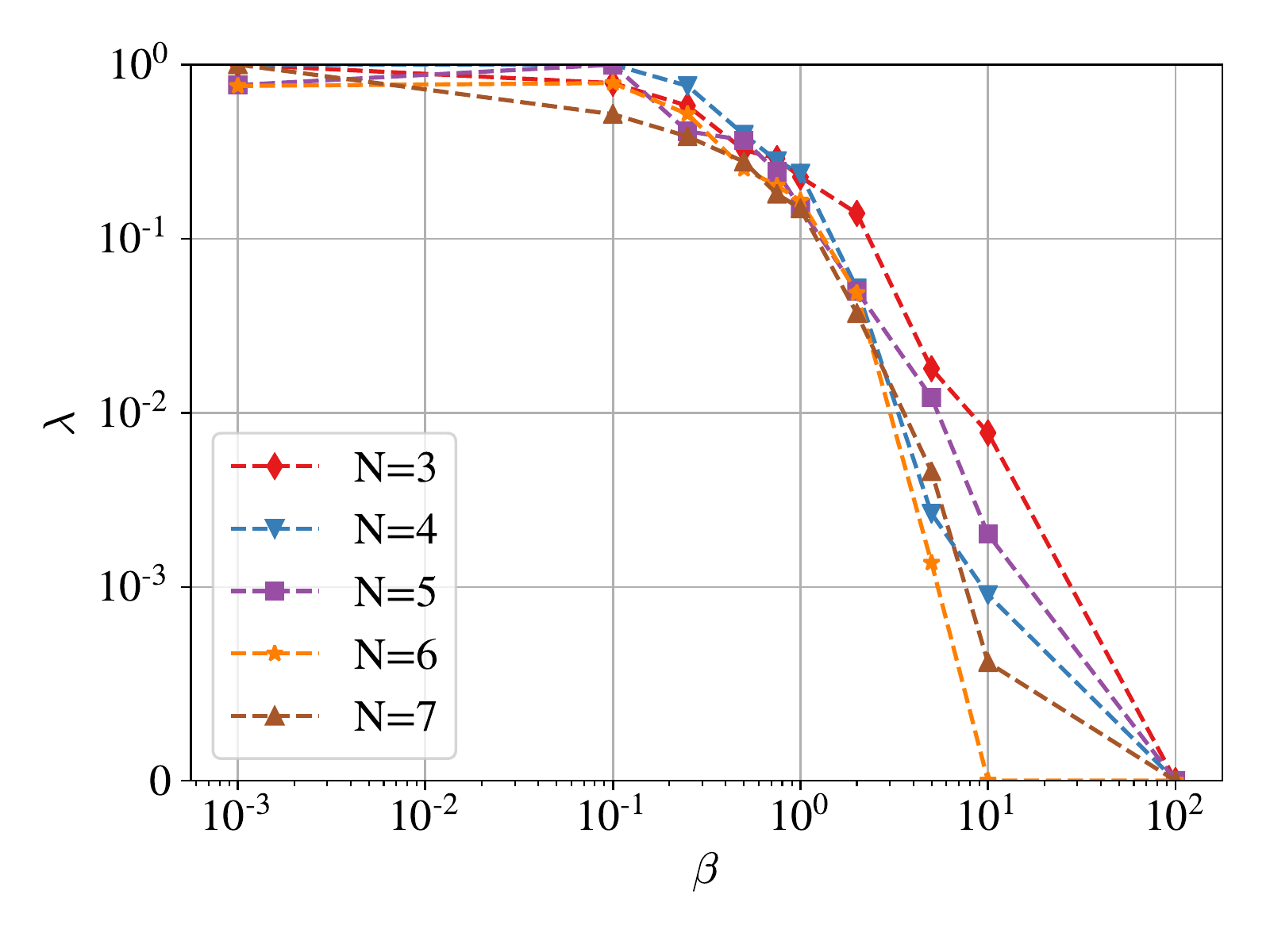}
    }\\
    \subfloat[Transverse-field Ising chain with uniform coefficients.\label{fig:beta-vs-lambda-tfi-u}]{    
        \includegraphics[width=0.49\linewidth]{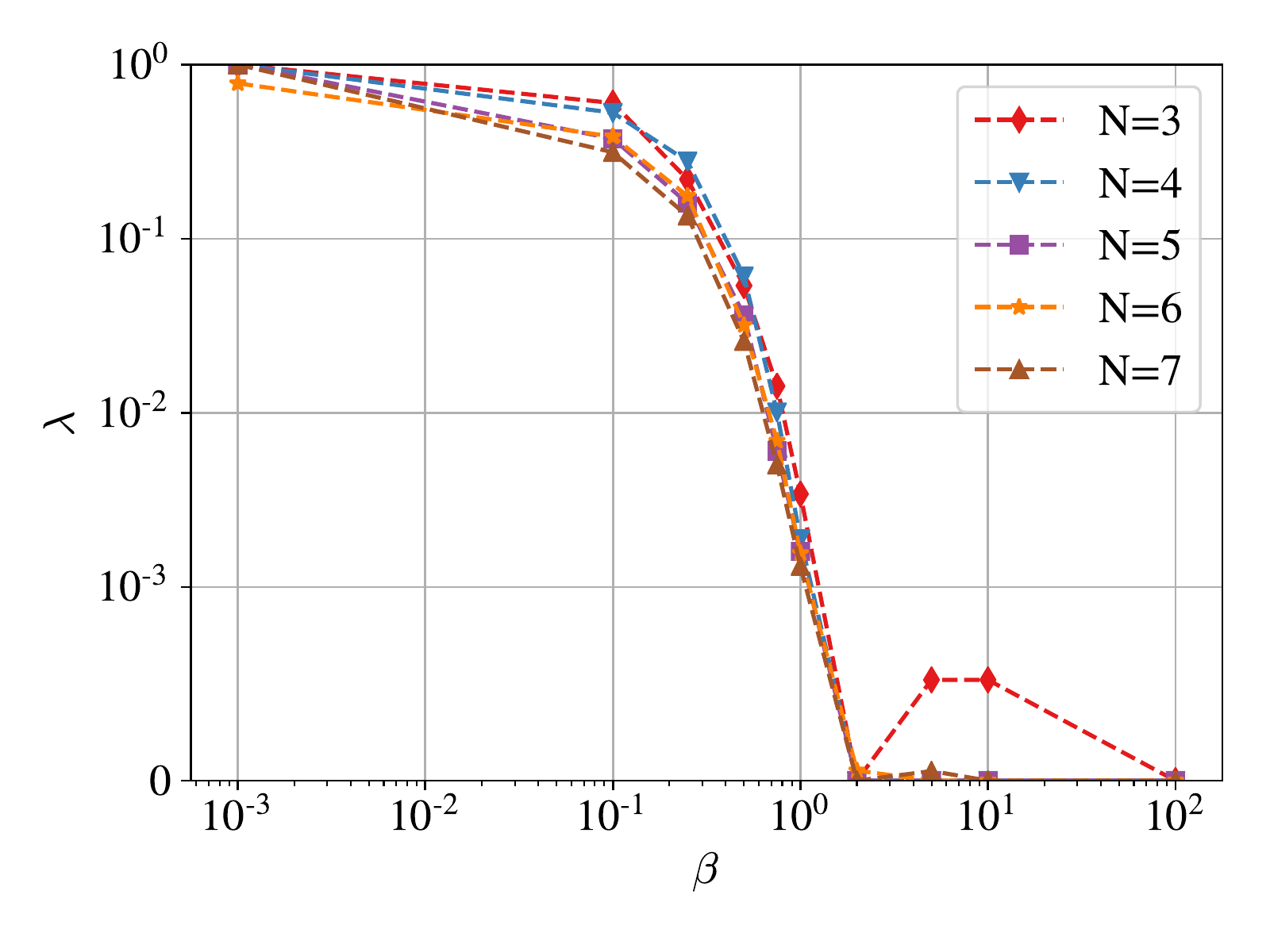}
    }
    \subfloat[Transverse-field Ising chain with random coefficients.\label{fig:beta-vs-lambda-tfi-r}]{
        \includegraphics[width=0.49\linewidth]{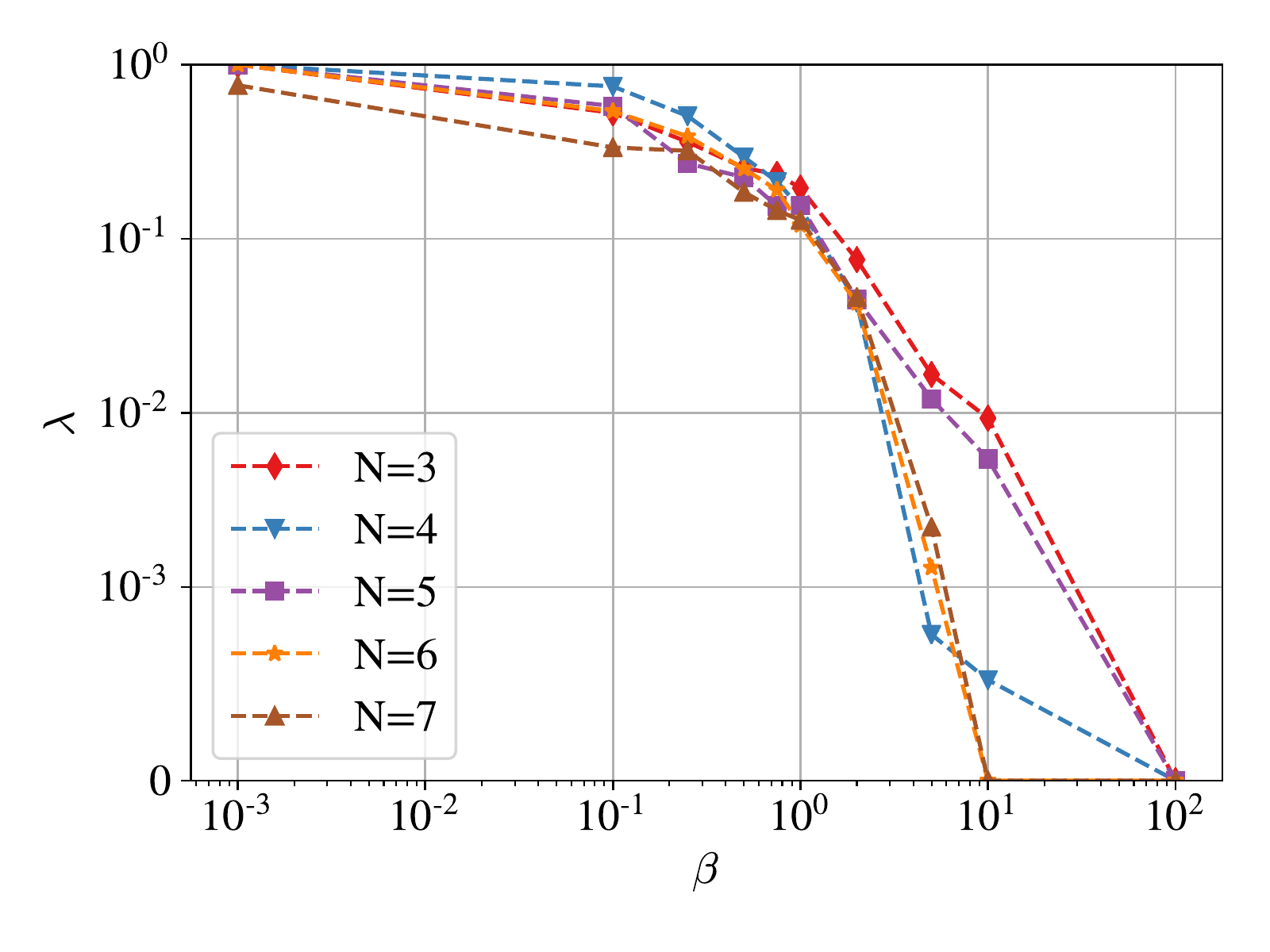}
    }\\
    \subfloat[1D Heisenberg model with uniform coefficients.\label{fig:beta-vs-lambda-Heisenberg-u}]{
        \includegraphics[width=0.49\linewidth]{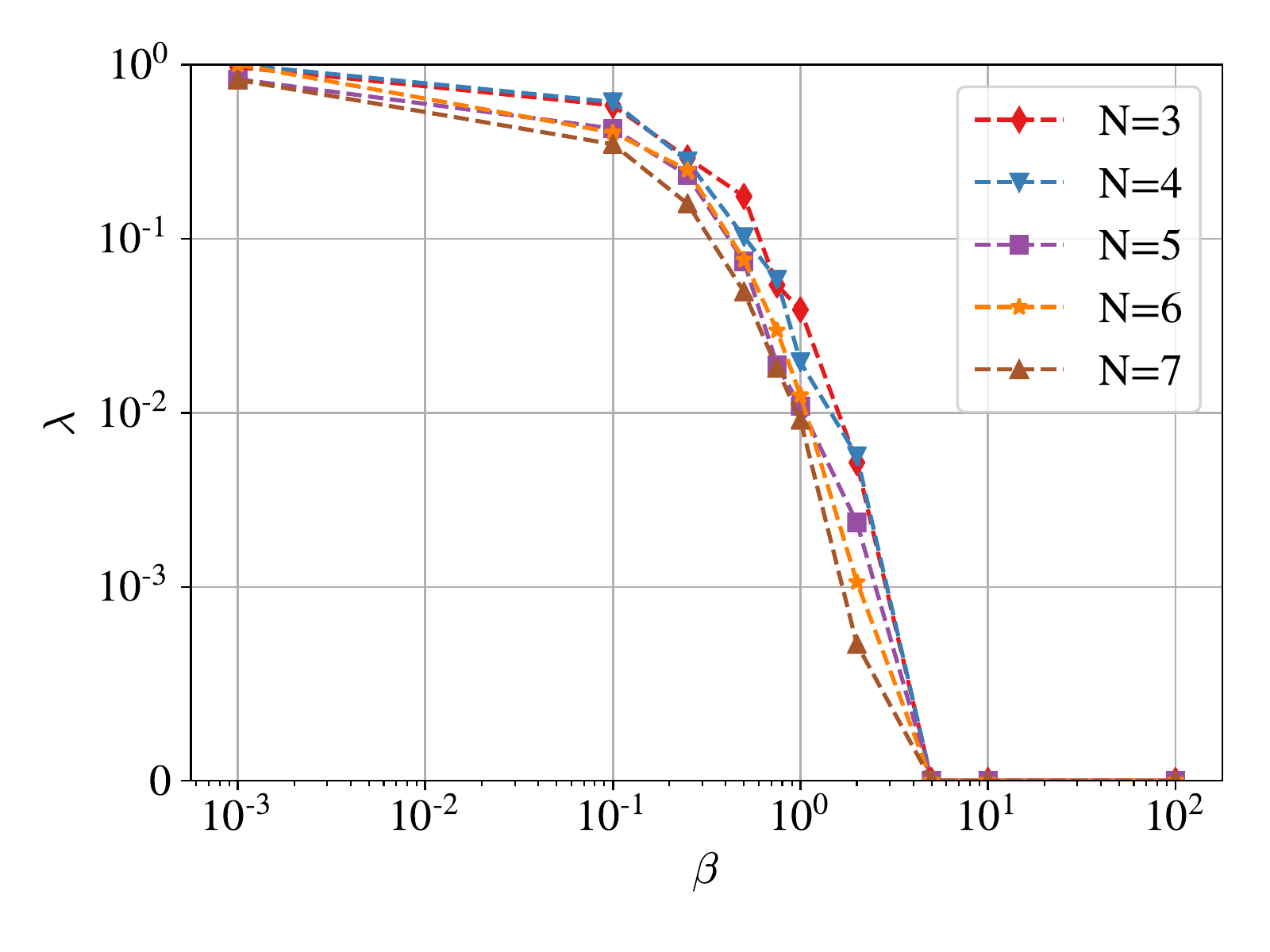}
    }
    \subfloat[1D Heisenberg model with random coefficients.\label{fig:beta-vs-lambda-Heisenberg-r}]{
        \includegraphics[width=0.49\linewidth]{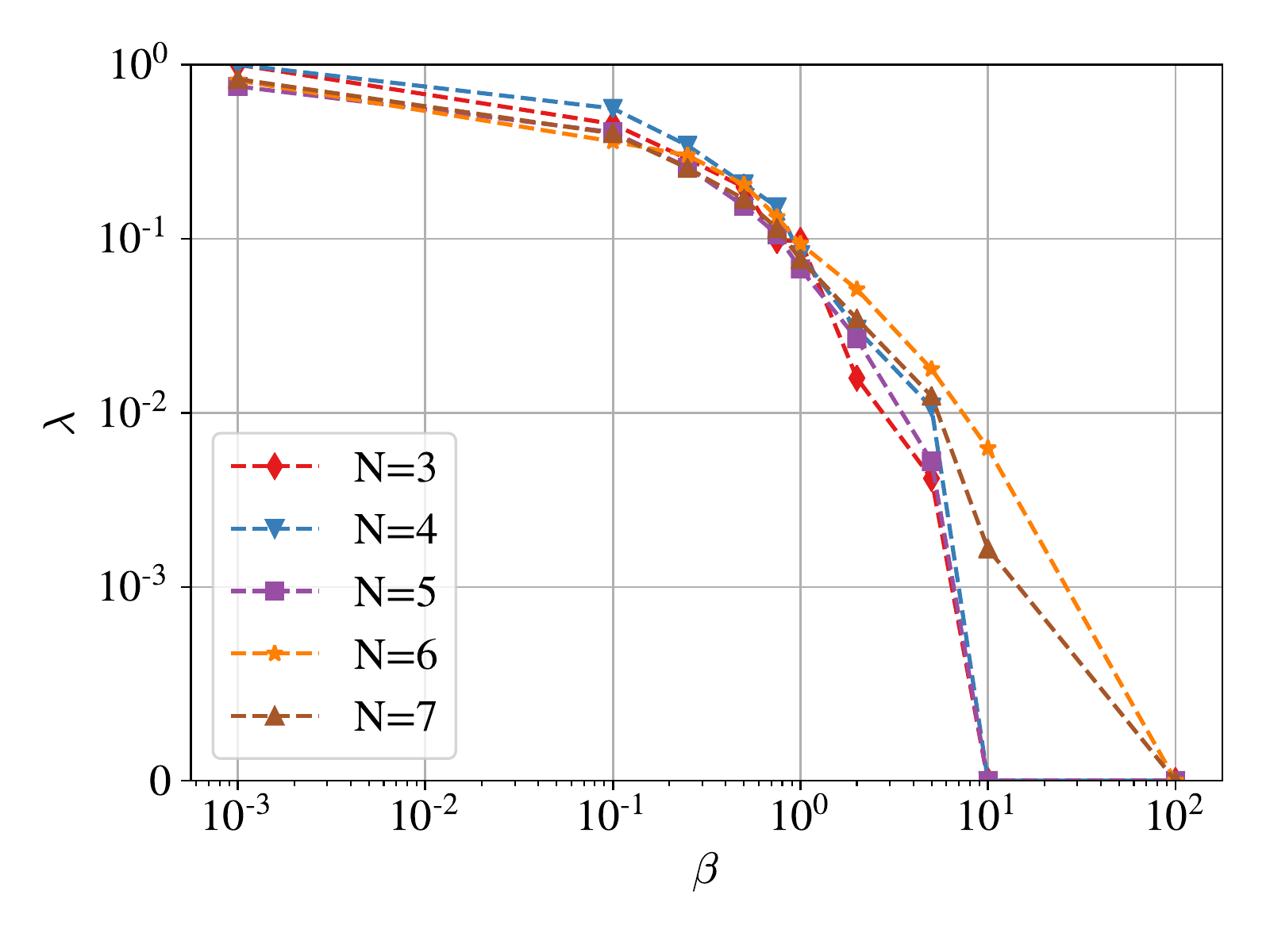}
    }
    \caption{Final noise level $\lambda$ as a function of the inverse temperature $\beta$ for various models and system sizes. We used a \textit{symlog} scale for the y-axis, hence the scale becomes linear below $10^{-3}$. We observe a clear decrease of the noise level with $\beta$, with $\lambda \approx 1$ for $\beta=10^{-3}$ (corresponding to the maximally-mixed state) and $\lambda \approx 0$ for $\beta \approx 10^2$ (corresponding to the ground-state). It shows that the general relationship between the noise and the temperature has overall been correctly learned by our model.\label{fig:beta-vs-lambda}} 
\end{figure*}

We first report the fidelity between the learned state and the thermal state as a function of the inverse temperature $\beta$ for all the different models in \cref{fig:beta-vs-F}. 
We also report the final noise level $\lambda$ as a function of $\beta$ for those models in \cref{fig:beta-vs-lambda}. 
Finally, we present the optimization curves for all the models at $N=4$, for three different temperatures $\beta \in \{ 0.1, 0.5, 10 \}$, in \cref{fig:learning-curves}. We can notice a few phenomena from those figures:
\begin{enumerate}
    \item The optimization curves presented in \cref{fig:learning-curves} show that the optimization procedure improves the solution compared to a random initialization, both when a very high fidelity is reached at the end and when the fidelity is lower. It eliminates the possibility that random states being closed to the desired thermal states would explain our results. Moreover, the fidelity tends to constantly increase with the number of iterations, showing that our approximate cost-function might be well-suited to our optimization goal.
    \item Thermal states at low and high temperatures are easily approximated by our method, for all models and system sizes. Looking at the $\lambda$ curves, we see that the optimizer is indeed able to find $\lambda=0$ for very large $\beta$ and $\lambda=1$ for very low $\beta$. Hence, when the thermal state gets close to a maximally-mixed state or to a pure state, the algorithm learns to respectively maximize or minimize the noise, independently of the initial noise level.
    \item The performance tends to degrade at intermediate temperatures, reaching for instance a fidelity of $0.6$ for the Heisenberg model with random coefficients. However, there are several temperatures for which a non-trivial noise level is learned and the fidelity remains high, such as the same model at $\beta=10^{-1}$, for which a fidelity above $96\%$ is reached for all system sizes with a noise level between $0.5$ and $0.8$. Hence the algorithm can actually find the correct thermal state in non-trivial temperature regimes.
\end{enumerate}

From those results, an important question to consider is whether the low fidelity obtained for some systems is due to a failure of the optimization procedure or to the potentially low expressibility of our noisy ansatz. To tackle this question, we tested different methods to optimize the parameters of the ansatz, including a grid-search in the parameter space for systems that are small enough to allow it to run in a reasonable time. We found no significant improvement in the fidelity compared to the original optimization method. We also tried to initialize the unitary ansatz to the ground-state solution before turning on the noise, but it did not result in a significant increase of fidelity neither. Finally, to evaluate the effect of our free-energy approximation, we performed all the experiments previously mentioned using finite-difference on the true free-energy. The corresponding results can be found in Supplementary \ref{fig:true-free-energy-optimization}, where we observe very similar fidelities as with the approximate free-energy method. It means that for the hardest systems tested in this work, the noisy ansatz was probably not expressible enough to output an accurate approximation of the thermal state, independently of the optimization algorithm. Changing the depolarizing gates to more general noise channels could help improve the expressibility of the ansatz and is let for future work.

%% file: sections/6discussion.tex
\section{Discussion} \label{sec:discussion}
%The Discussion should be succinct and must not contain subheadings.
In this paper, we introduced a novel type of variational algorithms, in which the noise is parameterized and optimized together with the unitary gates. We used this architecture to prepare thermal states, overcoming some of the most common challenges for this task, such as the need of ancilla qubits and the adverse effect of noise. To optimize our ansatz, we used a closed-form approximation of the free-energy and performed gradient-descent with it. We investigated various Hamiltonians and deduced that the ability of our method to learn the correct thermal states strongly depends on the model, the temperature and the system size. While we systematically obtained fidelities above $0.9$ for both the transverse-field and the classical Ising chain, we had fidelities below $0.7$ at some temperatures for the 1D Heisenberg model with random coefficients. We also identified a specific range of temperatures for each model, for which the task is harder for NAVQT to solve. Our experiments with different optimization algorithms reveal that the failure of the ansatz to learn the correct thermal state in those cases is probably an expressibility rather than an optimization issue. 

This paper serves as a starting point in the study of noise-assisted thermalization, and many avenues are still open for future work. For instance, we only considered a single shared parameter $\lambda$ for all the depolarizing gates, as it allowed us to derive an approximation of the free energy, which simplified the optimization process. Varying the noise across each layer and each qubit independently could significantly increase the expressibility of the ansatz. More generally, including dephasing channels or asymmetric depolarizing gates in the ansatz could also result in a more expressible circuit. However, new methods for obtaining and minimizing the free energy would be needed in those cases.

A second important aspect for future work would be to better understand the theory behind noise-assisted variational circuits. For instance, what are the conditions on the Hamiltonian and the temperature under which NAVQT can approximate the thermal state with an arbitrary high fidelity? How does our method scale with the system size? What type of noise is necessary to approximate a given thermal state? 

Finally, it could be interesting to study the optimization landscape of NAVQT and potentially come up with optimization algorithms that are more tailored to this problem. For instance, it has been shown that a barren plateau phenomenon occurs in noisy circuits that are similar to our ansatz~\cite{wang2020noise}. It can potentially hinder the scalability of our method, as it relies explicitely on increasing the noise. Finding the relationship between the temperature $\beta$, the system size $N$ and the magnitude of the gradient could be an interesting direction for future research.

%% file: sections/acknowledgements.tex
\section*{Acknowledgements} \label{sec:acknowledgements}
We would like to thank Michael Kastoryano, Jonatan Bohr Brask and Daniel Stilck França for fruitful discussions during this work, as well as Dan Browne for useful discussions and feedback during the writing of this manuscript. We would also like to thank Guillaume Verdon and Antonio Martinez for providing helpful tutorials and advice on noisy simulations of variational quantum circuits with Tensorflow Quantum. AP was supported by the Engineering and Physical Sciences Research Council [grant number EP/S021582/1]. JF was supported by the William Demant Foundation [grant number 18-4438]
%We would like to thank Michael Kastoryano for fruitful discussions in the initial phase of this work, as well as Jonatan Bohr Brask, Daniel Stilck França and Dan Browne for useful discussions and feedback during the writing of this manuscript. We would also like to thank Guillaume Verdon and Antonio Martinez for providing helpful tutorials and discussions on noisy simulations of variational quantum circuits with Tensorflow Quantum. AP was supported by the Engineering and Physical Sciences Research Council [grant number EP/S021582/1].

\twocolumngrid

%% file: sections/appendix.tex
\subsection*{Supplementary Note: Estimating the free energy of a noisy circuit}

In order to learn the thermal state with NAVQT, we need to minimize the free energy. Obtaining the free energy from the output of a quantum circuit is hard, since the entropy is a highly non-linear function of the state. For unitary evolutions, the entropy is constant, but for non-unitary circuits, including our ansatz, the entropy needs to be estimated for each change of parameters. To simplify this task, we consider the following approximation, represented in Fig. 1c of the main manuscript: all the depolarizing gates are shifted to the beginning of the circuit. Using this circuit, it is now possible to compute the entropy analytically.

If $m$ is the number of layers of our unitary ansatz, the approximate circuit consists in the composition of $m$ depolarizing gates $\D(\lambda)$ for each qubit. We can now use the fact that the composition of depolarizing gates is itself a depolarizing gate:
\begin{equation}
    \D(\lambda_2) \circ  \D(\lambda_1) = \D(1-(1-\lambda_{2})(1-\lambda_{1}))
\end{equation}
or more generally
\begin{equation}
    \D(\lambda_{m}) \circ ... \circ \D(\lambda_1) = \D(1-(1-\lambda_{m})...(1-\lambda_{1}))
\end{equation}

Assuming that the noise parameter $\lambda$ is the same for all the gates, then the above simplifies to $\D(1-(1-\lambda)^{m})$. If we note $\Lambda = 1-(1-\lambda)^{m}$ this new parameter, the entropy of $m$ consecutive noise gates acting on a single qubit initialized with $\ket{0}$ can be written as
\begin{equation}
    \begin{split}
        S(\rho_{\Lambda}) &= - \Tr[\rho_{\Lambda} \ln (\rho_{\Lambda})]\\
        &= - \Tr[\left((1-\Lambda) \ketbra{0} + \Lambda\frac{\One}{d}\right) \ln \left((1-\Lambda) \ketbra{0} + \Lambda\frac{\One}{d}\right)]\\
        &= - \bigg[\left((1-\Lambda) + \frac{\Lambda}{d} \right) \ln ((1-\Lambda) + \frac{\Lambda}{d}) +  \frac{(d-1)\Lambda}{d} \ln(\frac{\Lambda}{d})\bigg] 
    \end{split}
\end{equation}
and substituting for $\Lambda = 1-(1-\lambda)^{m}$
\begin{equation}
    \begin{split}
     S(\rho_{\lambda}) &= - \bigg[\left((1-\lambda)^{m} + \frac{(1-(1-\lambda)^{m})}{d} \right)  \ln ((1-\lambda)^{m} + \frac{(1-(1-\lambda)^{m})}{d}) \\
        &+  \frac{(d-1)(1-(1-\lambda)^{m})}{d} \ln(\frac{(1-(1-\lambda)^{m})}{d})\bigg] 
    \end{split}
\end{equation}
We now use the fact the entropy of a product state is the sum of the individual entropies to write
\begin{equation}
    S(\rho_{\lambda}^{\otimes N}) = N S(\rho_\lambda)
\end{equation}
which directly gives us the entropy of the state preceding the unitary ansatz. Since applying a unitary operation to a state does not change its entropy, it means that the overall entropy of the output state, that we call $S(\lambda)$, is given by the expression above. To optimize over it, we need to compute its gradient, which can also be obtained analytically as
\begin{equation}
    \begin{split}
        \nabla_{\lambda} S(\lambda) &= N \frac{d-1}{d} m (1 - \lambda)^{m-1} \left[-\ln(\frac{(1 - (1 - \lambda)^m)}{d}) + \ln(\frac{(1 - (1 - \lambda)^m + d (1 - \lambda)^m)}{d})\right]
    \end{split}
\end{equation}

The overall free energy, which we want to minimize, is given by
\begin{equation}
    F(\thetab, \lambda) = E(\thetab, \lambda) - T S(\lambda).
\end{equation}
The gradient of the energy with respect to $\theta$ can be efficiently computed on a quantum device using the parameter shift-rule, while its gradient with respect to $\lambda$ can be computed using finite-difference (since $E(\lambda)$ itself can easily be extracted from the output of the circuit). Therefore, the overall gradient, given by
\begin{align}
    \nabla_{\thetab} F(\thetab,\lambda) &= \nabla_{\thetab} E(\thetab,\lambda) \\
    \nabla_{\lambda} F(\thetab,\lambda) &=  \nabla_\lambda E(\thetab,\lambda) - T \nabla_\lambda S(\lambda)
\end{align}
can be efficiently computed using our circuit approximation.

\subsection*{Supplementary Figures}
\begin{figure}[H]
    \centering
    \subfloat[Entropy comparison]{
        \includegraphics[width=0.49\textwidth]{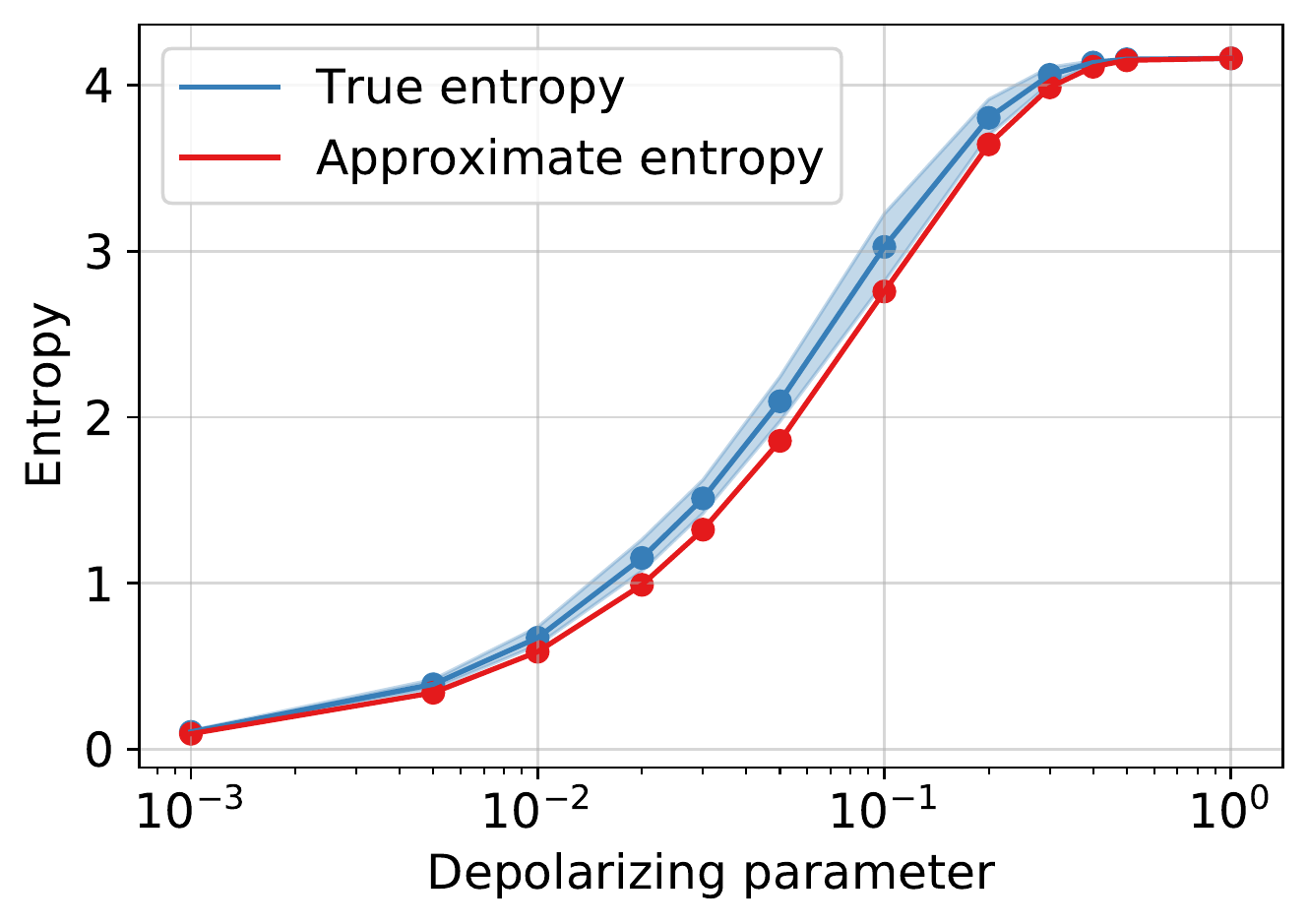}
    }
    \subfloat[Energy comparison]{
        \includegraphics[width=0.49\textwidth]{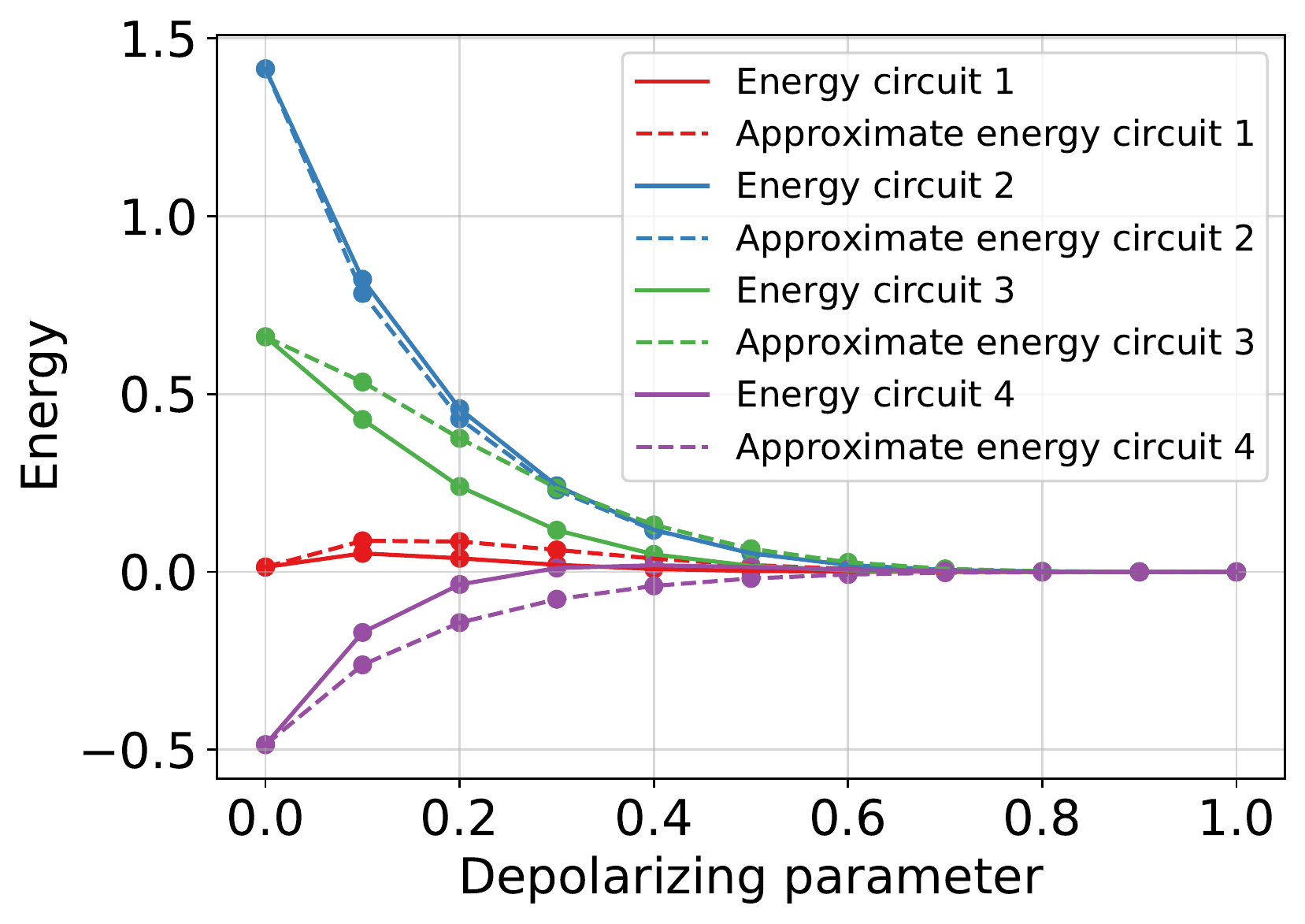}
    }
    \caption{Comparison of the approximate ansatz with the true one, for both the entropy and the energy as a function of the depolarizing noise, for random circuits with $6$ qubits and $3$ layers. \textbf{(a)} Entropy of the two circuit types. Since the entropy of the true circuit depends on the unitary parameters, we sampled $100$ random parameters and took the average, minimum and maximum of the entropy (blue area). We see that the two curves follow a similar trajectory, with the approximate entropy being a lower bound on the true one. 
    \textbf{(b)} Energy of the two circuit types for the transverse-field Ising model with uniform coefficients. Each color represents a circuit with different random unitary parameters. We see that the approximate energy tends to be close to the true one, following an overall similar trajectory.
    \label{fig:entropy_energy_simulations}}
\end{figure}

\begin{figure}[b!]
    \centering
    \subfloat[IC with uniform coefficients.]{
        \includegraphics[width=0.45\linewidth]{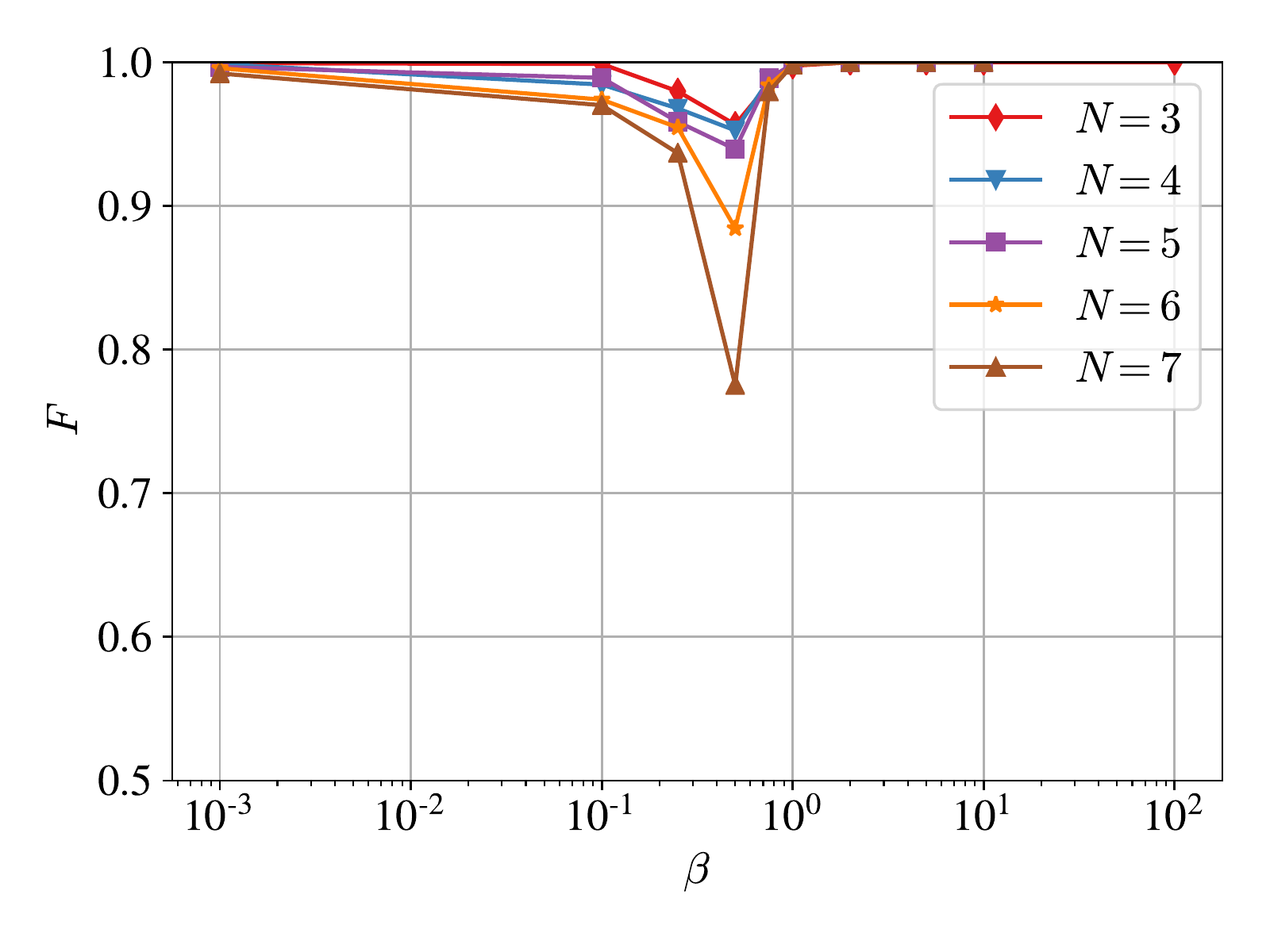}
    }
    \subfloat[IC with random coefficients.]{
        \includegraphics[width=0.45\linewidth]{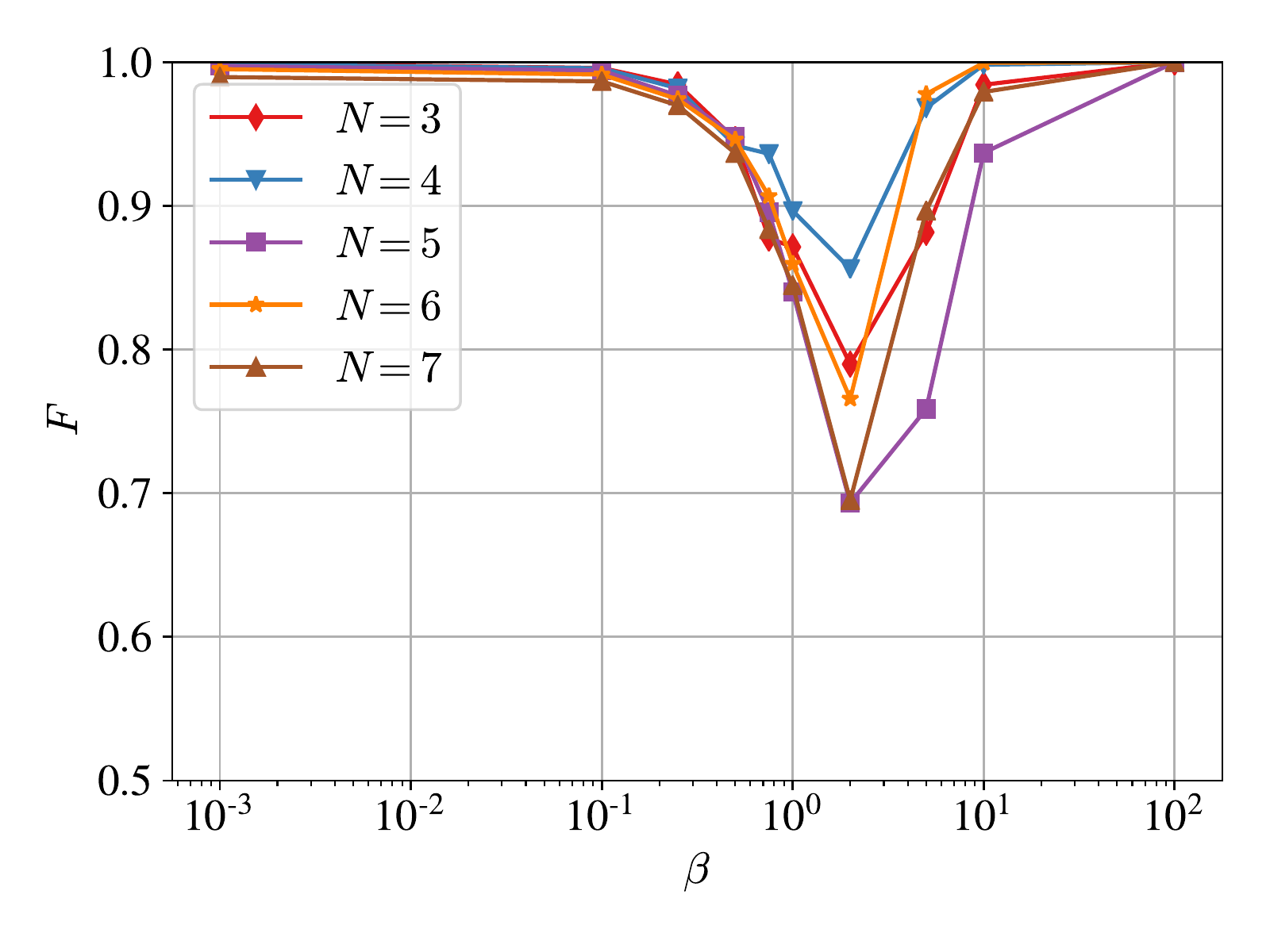}
    }\\
    \subfloat[TFI with uniform coefficients.]{
        \includegraphics[width=0.45\linewidth]{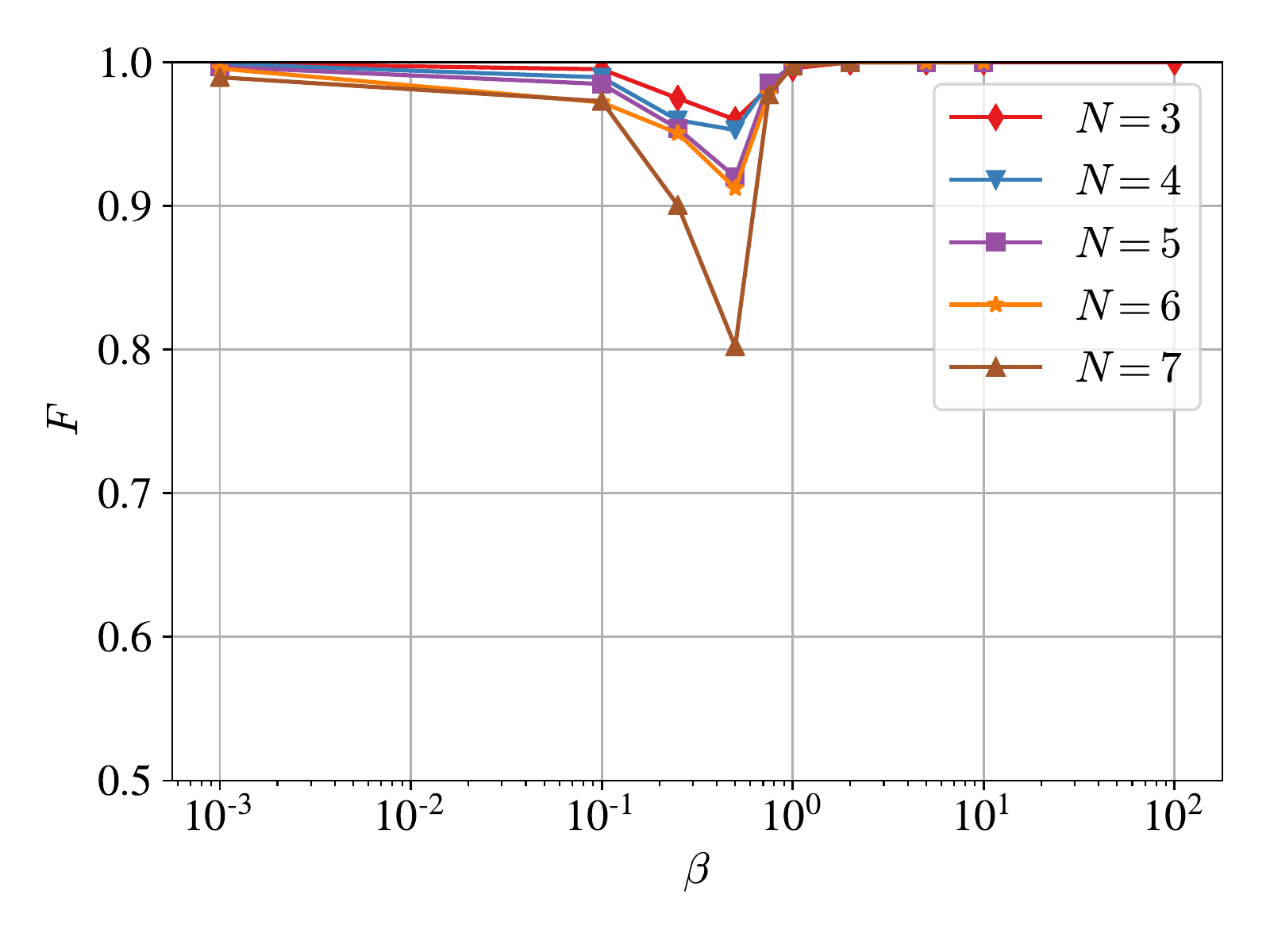}
    }
    \subfloat[TFI with random coefficients.]{
        \includegraphics[width=0.45\linewidth]{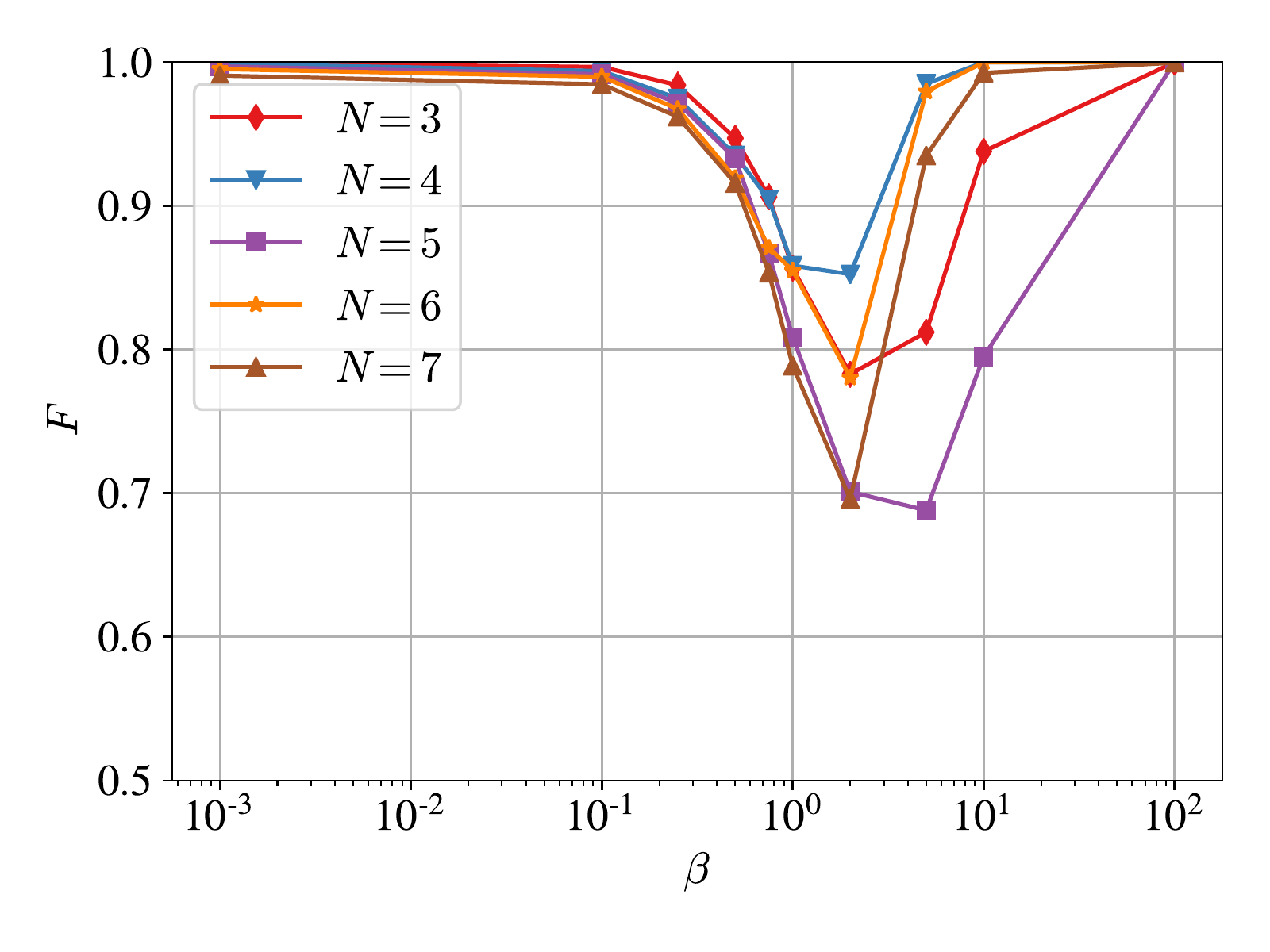}
    }\\
    \subfloat[Heisenberg with uniform coefficients.]{
        \includegraphics[width=0.45\linewidth]{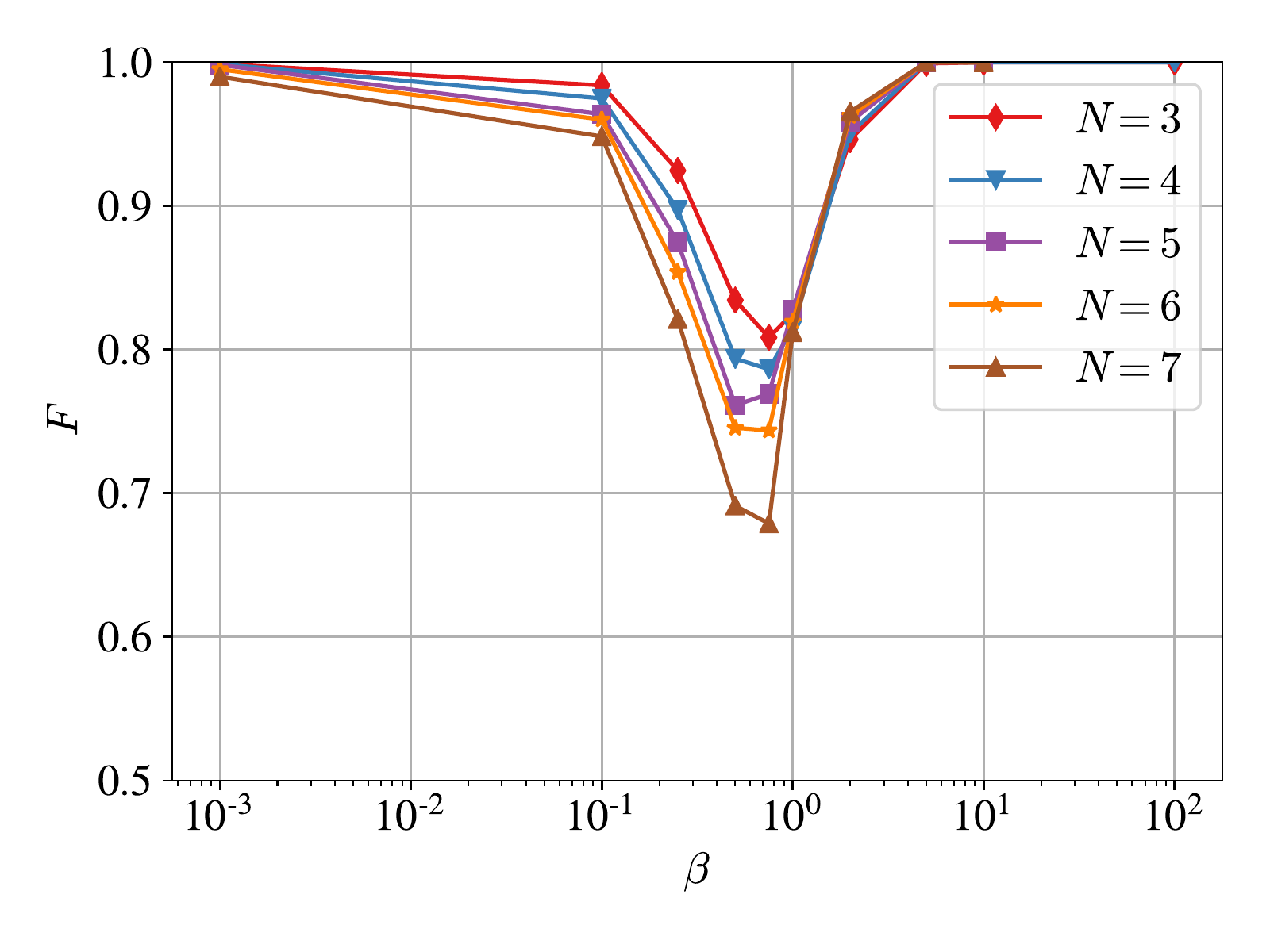}
    }
    \subfloat[Heisenberg with random coefficients.]{
        \includegraphics[width=0.45\linewidth]{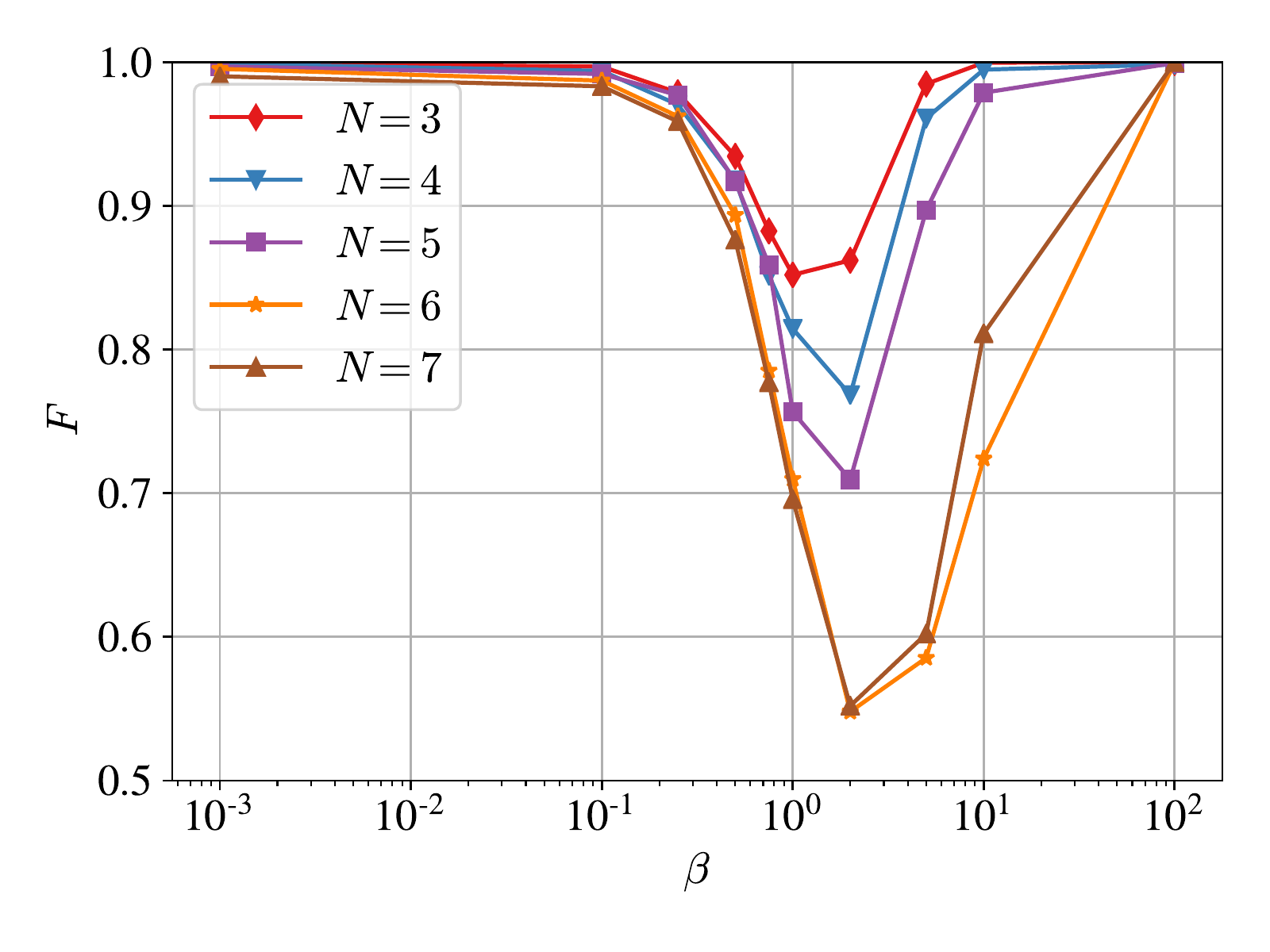}
    }

    \caption{Fidelities obtained when minimizing the actual free energy (as opposed to the approximation), using finite-difference to obtain the gradient. We see that the results are similar to those in Fig. 3 of the main manuscript, where the approximate free energy is used for optimization. It shows that the approximation serves as a good heuristics for optimizing our NAVQT ansatz.
    \label{fig:true-free-energy-optimization}
    }
\end{figure}